# Thermodynamics and kinetics of binary nucleation in ideal-gas mixtures

Nikolay V. Alekseechkin

Akhiezer Institute for Theoretical Physics, National Science Centre "Kharkov Institute of Physics and Technology", Akademicheskaya Street 1, Kharkov 61108, Ukraine
Email: n.alex@kipt.kharkov.ua

**Abstract**. The nonisothermal single-component theory of droplet nucleation (Alekseechkin, 2014) is extended to binary case; the droplet volume $V$ , composition $x$, and temperature $T$ are the variables of the theory. An approach based on macroscopic kinetics (in contrast to the standard microscopic model of nucleation operating with the probabilities of monomer attachment and detachment) is developed for the droplet evolution and results in the derived droplet motion equations in the space $(V, x, T)$ - equations for $\dot{V} \equiv dV/dt$, $\dot{x}$, and $\dot{T}$. The work $W(V, x, T)$ of the droplet formation is calculated; it is obtained in the vicinity of the saddle point as a quadratic form with diagonal matrix. Also the problem of generalizing the single-component Kelvin equation for the equilibrium vapor pressure to binary case is solved; it is presented here as a problem of integrability of a Pfaffian equation. The equation for $\dot{T}$ is shown to be the first law of thermodynamics for the droplet, which is a consequence of Onsager's reciprocal relations and the linked-fluxes concept. As an example of ideal solution for demonstrative numerical calculations, the *o*-xylene-*m*-xylene system is employed. Both nonisothermal and enrichment effects are shown to exist; the mean steady-state overheat of droplets and their mean steady-state enrichment are calculated with the help of the $3D$ distribution function. The qualitative peculiarities of the nucleation thermodynamics and kinetics in the water-sulfuric acid system are considered in the model of regular solution. It is shown that there is a small kinetic parameter in the theory due to the small amount of the acid in the vapor and, as a consequence, the nucleation process is isothermal.

## 1. Introduction

Despite decades of development, the theory of binary nucleation is still attractive and topical. One of the challenges to the accuracy of the theory is non-ideality of a new phase solution, since there is no exact analytical dependence of the chemical potential of a component of such solution on the composition, $\mu_i(x)$. Non-ideality also leads to the specific effect revealed in the past [1, 2] by the numerical solution of thermodynamic equations: in a vapor mixture of two substances, e.g. water and sulfuric acid, nucleation arises even at partial pressures of these substances less than their saturation pressures at the given temperature (i.e. when both the substances are “undersaturated” for the formation of droplets of pure components); furthermore, nucleation occurs at an extremely small amount of sulfuric acid in the vapor. Of course, exact analytical description of this effect is problematic due to the lack of the exact dependence $\mu_i(x)$. However, a qualitative explanation based on the simple model of regular solution can be done, which is demonstrated in the present report.

Systematic theoretical study of binary nucleation starts from the classical work by Reiss [3], where the single-component kinetic theory by Zeldovich and Frenkel [4, 5] was generalized to the binary case. The main features of the model used in this paper are as follows. The new phase embryo (the droplet of solution) is described by the numbers $N_1$ and $N_2$ of monomers constituting it; the probabilities of attachment and detachment of each kind monomers, $w_i^{(+)}$ and $w_i^{(-)}$, are introduced and the flux of embryos in the two-dimensional $(N_1, N_2)$-space is considered. The nucleation work $W(N_1, N_2)$ is the saddle surface in this space, so that its normal section in the direction of the flux of embryos is the nucleation barrier; the direction itself is chosen by Reiss as the steepest descent one, i.e. it is determined by thermodynamics only. More accurate later studies [6-8] showed that kinetics also “participates” in the selection of this direction – the diffusivities in kinetic equations deflect the flux of nuclei from the steepest descent, so that the flux direction is self-consistently determined by both thermodynamic and kinetic parameters.

The method of the work [7] was subsequently extended [9] to an arbitrary number of variables $\{X_i\}$ (one of them, $X_1$, is unstable) and the following equation for the steady state nucleation rate was obtained:

$$I = C_0\sqrt{\frac{kT}{2\pi}\left|h_{11}^{-1}\right|}\left|\kappa_1\right| e^{-\frac{W_*}{kT}} \tag{1}$$

where $T$ is the temperature of the system; $W_*$ is the nucleation work of the critical nucleus. The work $W(X_1,\dots,X_n)$ in the vicinity of the saddle point $\{X_i^*\}$ has the form

$$W(X_1,\dots,X_n) = W_* + (1/2)\sum_{i,k=1}^{n} h_{ik}(X_i - X_i^*)(X_k - X_k^*) \tag{2}$$

The quantity $h_{11}^{-1}$ in Eq. (1) is the matrix $\mathbf{H}^{-1}$ element; $\kappa_1$ is the negative eigenvalue of the matrix $\mathbf{Z}$ of the nucleus motion equations near the saddle point

$$\dot{X}_i = -\sum_{k=1}^{n} z_{ik}(X_k - X_k^*), \quad \mathbf{Z} = \mathbf{DH}/kT, \tag{3}$$

where $\mathbf{D}$ is the matrix of “diffusivities” in the Fokker-Planck equation for the distribution function of nuclei $f(X_1,\dots,X_n;t)$. The quantity $C_0$ is the normalizing factor of the one-dimensional equilibrium distribution function $f_e(X_1)$; it is determined in the framework of a statistical mechanical approach [10-12].

The *microscopic* model with the probabilities $w_i^{(+)}$ and $w_i^{(-)}$ mentioned above seems to be natural in the processes of droplet nucleation; therefore, it has become basic for the works on binary nucleation. In addition, the finite-difference kinetic equations underlying it are convenient for numerical simulations of nucleation [13-14]. On the other hand, this model also has some shortcomings. First, there exist difficulties with normalizing the equilibrium distribution function $f_e(N_1, N_2)$, since the variables $N_1$ and $N_2$ are physically equivalent (there is no stable and unstable variables). The search for ways of normalizing this function led to the need to "break the symmetry" of these variables, i.e. to go to the new variables $N = N_1 + N_2$ and $x = N_1/(N_1 + N_2)$ [9, 15]. The roles of the variables $N$ and $x$ are quite different: $N$ is an unstable variable, whereas $x$ is a stable one. Thus the process of binary nucleation in these variables becomes a *process with linked fluxes* [16]. Accordingly, normalization of the function $f_e(N, x)$ with respect to $x$ is performed with the help of the fluctuation theory, whereas the result of the one-dimensional theory is employed for normalizing this function with respect to $N$. The function $f_e(N_1, N_2)$ is then obtained from the function $f_e(N, x)$ by the inverse transformation of variables.

Second, the given model does not allow investigating the "viscosity effect" in droplet nucleation [17], since it results in differential growth equations of the first order, whereas a second-order differential equation for an unstable variable is required for studying this effect. Third, this model is not universal for nucleation generally. For example, knowledge of the probabilities $w_i^{(+)}$ and $w_i^{(-)}$ is insufficient to describe the nucleation kinetics of vapor bubbles in a liquid. These probabilities determine the amount of vapor in the bubble, however, the bubble volume (unstable variable) can vary independently of the presence of vapor in it (in some limiting case, the vapor can be neglected at all [4]).

All these shortcomings are overcome, when the *macroscopic* approach is used, i.e. the approach based on macroscopic equations of motion of a nucleus in the space $\{X_i\}$, Eq. (3). Composition $x$ as a natural thermodynamic variable (the chemical potential depends on it) is one of the variables $\{X_i\}$; it is shown below that the matrix $\mathbf{H}$ has a diagonal form in this case, differently from the case, when the variables $(N_1, N_2)$ are used. Knowledge of the matrix $\mathbf{Z}$ allows calculating the nucleating rate (the quantity $\kappa_1$), the direction of the flux of droplets and their steady state distribution function. New advantages appear in the macroscopic approach as well, in particular, ease of studying the nonisothermal effect in nucleation [16, 18-24], as was demonstrated in the previous work [17]. In general, the mathematical formalism employed in the macroscopic approach is simpler than in the microscopic one. This approach was shown to be fruitful both in droplet and bubble nucleation [25, 26].

Two results of the previous work [17] on droplet nucleation in a single component vapor are employed here. (i) The limiting "high-viscosity" case holds in the ideal gas approximation, so that the differential equation for an unstable variable (the growth equation) is a first-order equation, i.e. equation for $\dot{V}$, $\dot{R}$, or $\dot{N}$ ($V$, $R$, and $N$ are respectively the nucleus volume, radius, and number of molecules). (ii) The diffusion coefficient in a gas is sufficiently large (the estimates are made therein), so that the droplet growth is interface controlled. It was shown that the usual growth of a droplet in pure vapor and the diffusion growth in the presence of an inert background gas lead to the same growth equation. For these reasons, the equation for $\dot{N}$ obtained in this work is employed here for each of the species:

$$\dot{N}_i = \frac{gV^{2/3}\beta_i u_i}{kT_0}\left[P_{0i} - \sqrt{\frac{T_0}{T}}P_{ei}(R,T,x)\right], \quad i = 1,2 \tag{4}$$

where $P_{0i}$ is the partial pressure of component $i$ in the vapor; $P_{ei}$ is the equilibrium pressure of component $i$ for the droplet of radius $R$, temperature $T$, and composition $x$; $T_0$ is the vapor temperature; $gV^{2/3} = 4\pi R^2$ is the droplet surface area, $g = 3^{2/3}(4\pi)^{1/3}$; $u_i = \sqrt{kT_0 / 2\pi m_i}$ is the mean velocity of the molecule (of the mass $m_i$) moving towards the droplet; $\beta_i$ is the condensation coefficient of the $i$-th species molecules which is the fraction of accommodating molecules from the incident flux (the fraction $(1-\beta_i)$ is reflected back to the vapor). So, the thermodynamic part of the problem, in addition to the standard task of determination of the matrix $\mathbf{H}$ in Eq. (2), also includes the determination of the quantities $P_{ei}$ (or generalization of Kelvin's equation to a solution).

The paper is organized as follows. Section 2 is the thermodynamic part of the paper. Here the work of droplet formation and its second differential at the saddle point (the matrix $\mathbf{H}$) are calculated. Also the partial equilibrium pressures $P_{ei}$ included in Eq. (4) are determined. With the help of these pressures, different relations between the state parameters of the critical droplet and the vapor are further derived, in particular, the critical droplet radius for the given vapor state is determined. These relations are applied for ideal and regular solutions. The kinetic part of the problem (Section 3) includes writing the droplet motion equations and deriving the matrix $\mathbf{Z}$. The relation between the $(V,x,T)$- and $(N_1,N_2,E)$-theories is also shown here and kinetic limits within the $(V,x,T)$-theory are examined. The droplet flux direction in the $3D$ space and the steady state distribution function are calculated in Section 4. With the help of this function, the two nucleation effects – the mean overheat of droplets and their mean enrichment – are evaluated for two model systems. The summary of results is given in Section 5.

## 2. Thermodynamics of binary droplet formation

### 2.1. Work near the saddle point

The parameters of the vapor consisting of two components are denoted by the subscript 0: the temperature $T_0$, the pressure $P_0$, and the first component fraction $x_0$. The droplet state is characterized by the temperature $T$, the pressure $P$, and the first component fraction $x \equiv x_1 = N_1 / N$, $N = N_1 + N_2$; the second component fraction is $x_2 = 1 - x$. Let $G(T, P, N_1, N_2)$ be some extensive quantity of the droplet. Below its specific (per one molecule) quantity $g(T, P, x) = G / N$ and the partial specific quantities

$$g_i(T, P, x) = \left(\frac{\partial G}{\partial N_i}\right)_{T,P,N_{j\neq i}} , \quad G = \sum_{i=1}^{2} g_i N_i , \quad g = \sum_{i=1}^{2} g_i x_i \tag{5}$$

are used. As the quantity $G$, the volume $V$ and the entropy $S$ of the droplet are used here:

$$S = \sum_{i=1}^{2} s_i N_i = sN , \quad V = \sum_{i=1}^{2} \upsilon_i N_i = \upsilon N , \quad s = xs_1 + (1-x)s_2 , \quad \upsilon = x\upsilon_1 + (1-x)\upsilon_2 \tag{6}$$

The chemical potentials of components in the droplet and in the vapor are $\mu_i(T, P, x)$ and $\mu_{0i}(T_0, P_0, x_0)$, respectively. The conditions of equilibrium of the critical droplet (its parameters are denoted by asterisk) with the vapor are

$$T_* = T_0 \tag{7a}$$

$$P_* - P_0 = \frac{2\sigma}{R_*} \equiv P_L^* \tag{7b}$$

$$\mu_i(T_*, P_*, x_*) = \mu_{0i}(T_0, P_0, x_0), \quad i = 1, 2 \tag{7c}$$

Just as the equation $\mu(T_0, P_0 + 2\sigma / R) = \mu_0(T_0, P_0)$ for a single component vapor determines the dependence $P_0(R)$ (Kelvin's equation) and from it the critical radius $R_*$ for given $T_0$ and $P_0$, the system of Eqs. (7c) with account for Eqs. (7a) and (7b) determines the critical radius $R_*$ and the critical droplet composition $x_*$ for the given $T_0$, $P_0$, and $x_0$.

The work of forming an embryo in a multicomponent system and its second differential at the saddle point are [25, 27]

$$W = \sum_{i=1`}^{2} (\mu_i - \mu_{0i})(N_i + \overline{N}_i) + (T - T_0)(S + \overline{S}) - (P - P_0)V + \sigma A \tag{8}$$

where $\overline{N}_i$ and $\overline{S}_i$ are the one-sided superficial quantities: the number of $i$-th species particles and entropy of the cluster, respectively;

$$\left(d^2 W\right)_* = -\frac{P_L^*}{3V_*}(dV)^2 + \left\{\sum_{i=1}^{2} d\mu_i dN_i + dTdS - dPdV\right\}_* \tag{9}$$

where $(P_L^* / 3V_*) = -(2/9) g\sigma V_*^{-4/3}$.

From the following equations

$$dS = \sum_{i=1}^{2} \left(N_i ds_i + s_i dN_i\right), \quad dV = \sum_{i=1}^{2} \left(N_i d\upsilon_i + \upsilon_i dN_i\right),$$

$$d\mu_i = -s_i dT + \upsilon_i dP + \mu_i' dx , \quad \mu_i' \equiv \left(\frac{\partial \mu_i}{\partial x}\right)_{T,P} \tag{10}$$

one obtains

$$\sum_{i=1}^{2} d\mu_i dN_i + dTdS - dPdV = \sum_{i=1}^{2} \{N_i [ds_i dT - d\upsilon_i dP] + \mu_i' dN_i dx\} \tag{11}$$

Thus the unstable (extensive) variable $V$ is naturally determined by Eq. (9). Hence, the positive definite quadratic form given by Eq. (11) is a function of stable (intensive) variables $T$, $P$, and $x$; the quantities $s_i(T,P,x)$ and $\upsilon_i(T,P,x)$ are functions of these variables.

In order to transform the last addend in Eq. (11), we note that

$$dx = (1-x)\frac{dN_1}{N} - x\frac{dN_2}{N} \tag{12}$$

Further,

$$\sum_{i=1}^{2} \mu_i' dN_i dx = (d\mu_1)_{T,P} dN_1 + (d\mu_2)_{T,P} dN_2 \tag{13}$$

The isothermal-isobaric Gibbs-Duhem equation reads:

$$x d\mu_1 + (1-x) d\mu_2 = 0, \quad d\mu_2 = -\frac{x}{1-x} d\mu_1 \tag{14}$$

Then

$$d\mu_1 dN_1 + d\mu_2 dN_2 = \frac{N}{1-x} d\mu_1 \left\{ (1-x)\frac{dN_1}{N} - x\frac{dN_2}{N} \right\},$$

and

$$\sum_{i=1}^{2} \mu_i' dN_i dx = \frac{N}{1-x} \mu_1' (dx)^2 \tag{15}$$

In order to transform the expression in square brackets in Eq. (11), we employ the generalized Gibbs-Duhem equation for the extensive quantity $G$,

$$\sum_{i=1}^{n} N_i dg_i = \left(\frac{\partial G}{\partial T}\right)_{P,N_j} dT + \left(\frac{\partial G}{\partial P}\right)_{T,N_j} dP \tag{16}$$

If $G$ is Gibbs' free energy, then $g_i = \mu_i$ and Eq. (16) is the usual Gibbs-Duhem equation

$$\sum_{i=1}^{n} N_i d\mu_i = -SdT + VdP \tag{17}$$

Substituting $G = S$ and $G = V$ in Eq. (16) and employing the familiar thermodynamic relations $(\partial S/\partial T)_P = C_P/T$ and $(\partial S/\partial P)_T = -(\partial V/\partial T)_P$, where $C_P$ is the heat capacity of the droplet of composition $(N_1, N_2)$ at constant pressure, we have

$$\sum_{i=1}^{2} N_i ds_i = \frac{C_P}{T} dT - \left(\frac{\partial V}{\partial T}\right)_{P,N_j} dP \tag{18a}$$

$$\sum_{i=1}^{2} N_i d\upsilon_i = \left(\frac{\partial V}{\partial T}\right)_{P,N_j} dT + \left(\frac{\partial V}{\partial P}\right)_{T,N_j} dP \tag{18b}$$

It is easy to derive the following identities for the equation of state $P = P(V,T,x)$ of a binary solution:

$$\left(\frac{\partial V}{\partial T}\right)_{P,x}\left(\frac{\partial T}{\partial P}\right)_{V,x}\left(\frac{\partial P}{\partial V}\right)_{T,x} = -1 \tag{19a}$$

$$\left(\frac{\partial V}{\partial x}\right)_{P,T}\left(\frac{\partial x}{\partial P}\right)_{V,T}\left(\frac{\partial P}{\partial V}\right)_{T,x} = -1 \tag{19b}$$

This pair of equations generalizes the familiar single-component identity for the equation $P = P(V,T)$, Eq. (19a). Obviously, the condition of constant $N_j$ in Eqs. (18a) and (18b) results in the condition of constant $x$.

With the help of the relation

$$C_P - C_V = -T\frac{(\partial V/\partial T)_P^2}{(\partial V/\partial P)_T} \tag{20}$$

and Eq. (19a), one obtains from Eqs. (18a) and (18b) (the subscripts $N_j$ and $x$ are omitted for brevity)

$$\sum_{i=1}^{2}\{(N_i ds_i)dT - (N_i d\upsilon_i)dP\} = \frac{C_V}{T}(dT)^2 + H(dP,dT),$$

$$H(dP,dT) \equiv -\left(\frac{\partial V}{\partial P}\right)_T (dP)^2 - 2\left(\frac{\partial V}{\partial T}\right)_P dPdT + \left(\frac{\partial V}{\partial T}\right)_P\left(\frac{\partial P}{\partial T}\right)_V (dT)^2 \tag{21}$$

It is easy to check with the help of Eq. (19a) that the determinant of the matrix of the quadratic form $H(dP,dT)$ is identically equal to zero. Thus, allocating a complete square, we have

$$H(dP,dT) = -\left(\frac{\partial V}{\partial P}\right)_T\left[dP - \left(\frac{\partial P}{\partial T}\right)_V dT\right]^2 = -\left(\frac{\partial V}{\partial P}\right)_T\left[\left(\frac{\partial P}{\partial V}\right)_T dV\right]^2 = -N\left(\frac{\partial P}{\partial \upsilon}\right)_T (d\upsilon)^2 \tag{22}$$

(we remember that all derivatives are calculated at constant $N_j$ and, consequently, at constant $N$). Here the transition from $P$ to $\upsilon$ has been done.

In summary, Eq. (9) acquires the form

$$\left(d^2W\right)_* = -\frac{P_L^*}{3V_*}(dV)^2 + N_*\left\{-\left(\frac{\partial P}{\partial \upsilon}\right)_{T,x}^* (d\upsilon)^2 + \frac{\mu'_{1*}}{1-x_*}(dx)^2 + \frac{c_V}{T_0}(dT)^2\right\} \tag{23}$$

It is seen from here that the required matrix $\mathbf{H}$ is a $4\times4$ diagonal matrix. Further we employ the usual approximation of incompressible liquid, $(\partial\upsilon/\partial P)_{T,x} = 0$, i.e. $h_{\upsilon\upsilon} = \infty$; the variance of the quantity $\upsilon$ is $\left\langle(\upsilon-\upsilon_*)^2\right\rangle = kT_0 h_{\upsilon\upsilon}^{-1} = 0$. Thus we have the thermodynamic limit with respect to $\upsilon$ [9], this variable drops out of consideration (the dimensionality is reduced by unity) and $\mathbf{H}$ is

$$\mathbf{H} = \begin{pmatrix} -\frac{P_L^*}{3V_*} & 0 & 0 \\ 0 & N_*\frac{\mu'_{1*}}{1-x_*} & 0 \\ 0 & 0 & \frac{C_V^*}{T_0} \end{pmatrix} \tag{24}$$

It is seen from Eq. (6) that the partial volumes $\upsilon_i$ do not depend on $P$ in the approximation of incompressible liquid; in Eq. (19b) $(\partial\upsilon/\partial P)_{T,x}=0$ and $(\partial x/\partial P)_{\upsilon,T}=0$, whereas $(\partial\upsilon/\partial x)_{P,T}$ is obviously finite.

Positiveness of the element $h_{xx}$ is provided by the thermodynamic condition of stability $\mu_1'>0$. The thermodynamic limit $h_{xx}\to\infty$ corresponds to transition to the single component $(V,T)$-theory; accordingly, it is realized for the pure first ($x_*\to 1$) or second ($x_*\to 0$) component. In the latter case, we have for a dilute solution $\mu_1(T,P,x)=\psi(T,P)+kT\ln x$, $\mu_{1*}'=kT_0/x_*\to\infty$ at $x_*\to 0$.

## 2.2. Equations of equilibrium for the critical binary droplet

### 2.2.1. The case of composition-independent surface tension

Considering Eq. (7c) (asterisk is omitted) in the differential form $d\mu_i=d\mu_{oi}$ at constant $T_0$ (i.e. supersaturation in the system is created by changing $P_0$ or $x_0$), we have

$$\upsilon_i d\left(P_0+P_L\right)+\mu_i' dx=\upsilon_0 dP_0+\frac{kT_0}{x_{0i}}dx_{0i} \tag{25}$$

where the compositional dependence of the ideal-gas chemical potential, $kT_0\ln x_{0i}$, was used. Further, we neglect by the first term in the left-hand side, in view of $\upsilon_i<<\upsilon_0$, and express the right-hand side via the partial pressures, $P_{01}=P_0x_0$, $P_{02}=P_0(1-x_0)$, employing the equation of state $\upsilon_0=kT_0/P_0$:

$$\frac{\mu_i'}{kT_0}dx+\frac{\upsilon_i}{kT_0}dP_L-d\ln P_{0i}=0 \tag{26}$$

So, we have a Pfaffian equation for the vector field $\mathbf{F}=(\mu_i'/kT_0)\mathbf{i}+(\upsilon_i/kT_0)\mathbf{j}-\mathbf{k}$ in the space $\mathbf{r}=(x,P_L,\ln P_{0i})$. The necessary and sufficient condition for integrability of a Pfaffian equation by *one relation* $P_{0i}=P_{0i}(x,P_L)$ is $(\mathbf{F}\,rot\mathbf{F})=0$. The particular case of this condition is $rot\mathbf{F}=0$, or $\mathbf{F}=gradU$, i.e. the field is potential. We have

$$rot\mathbf{F}=\frac{1}{kT_0}\left(-\frac{\partial\upsilon_i}{\partial\ln P_{0i}},\ \frac{\partial\mu_i'}{\partial\ln P_{0i}},\ \frac{\partial\upsilon_i}{\partial x}-\frac{\partial\mu_i'}{\partial P_L}\right) \tag{27}$$

In droplet nucleation, the condition $P_L^*>>P_0$ is satisfied; thus, we can assume that $\mu_i$ does not depend on $P_0$ and $\mu_i(P)=\mu_i(P_L)$ (neglect of $\upsilon_i dP_0$ in Eq. (25) can be also attributed to this approximation). So, $\partial\mu_i'/\partial\ln P_{0i}=0$; $\partial\upsilon_i/\partial\ln P_{0i}=0$ in view of incompressibility and

$$\frac{\partial\mu_i'}{\partial P_L}=\frac{\partial}{\partial x}\left(\frac{\partial\mu_i}{\partial P_L}\right)=\frac{\partial\upsilon_i}{\partial x}$$

Hence, the third component of rotation is also zero. As a result, $rot\mathbf{F} = 0$ and Eq. (26) has the form $dU = 0$; its solution is $U = const$.

Integrating Eq. (26) from the state ($R = \infty$, $x_i = 1$) to the current state ($R$, $x$), we get

$$kT_0 \ln \frac{P_{0i}}{P_{si}(T_0)} = P_L + \Delta\mu_i, \quad \Delta\mu_i \equiv \mu_i(T_0, P, x) - \tilde{\mu}_i(T_0, P), \quad P_{si}(T_0) = c_i\, e^{-\frac{q_i}{kT_0}} \tag{28}$$

where $\tilde{\mu}_i(T_0, P)$ is the chemical potential of the pure liquid $i$-th component; $P_{si}(T_0)$ is the equilibrium vapor pressure over a large amount of the pure liquid $i$; $q_i$ is the heat of vaporization per one molecule. From here,

$$P_{0i} = P_{si}(T_0) \exp\left\{ \frac{2v_i\sigma / R + \Delta\mu_i}{kT_0} \right\} = P_{si}(T_0) A_i(T_0, P, x) \exp\left( \frac{2v_i\sigma}{kT_0 R} \right) \tag{29}$$

where $A_i$ are the *activities* defined by relation $\Delta\mu_i = kT_0 \ln A_i$. For the flat interface, Eq. (29) gives the familiar definition of activities $A_i = P_{0i}^{(sol)} / P_{si}$ which is employed for their experimental determination; $P_{0i}^{(sol)}$ is the equilibrium pressure of the $i$-th species over a large amount of the solution of composition $x$.

Eq. (29) is a generalization of Kelvin's equation to solutions. It is important in the above analysis that the variables $P_L$ and $x$ are *independent*, which means that *the surface tension does not depend on* $x$.

### 2.2.2. The case of composition-dependent surface tension

The problem of thermodynamic consistency in the case of a composition-dependent surface tension was discussed in literature in past [28, 29]. Here it is formulated as a problem of integrability of a Pfaffian equation. When the surface tension depends on composition, we have $P_L(x) = 2\sigma(x)/R$ ($\sigma$ also may depend on $R$) and

$$dP_L = P'_{L,R} dR + P'_{L,x} dx, \quad P'_{L,R} \equiv \partial P_L / \partial R, \quad P'_{L,x} \equiv \partial P_L / \partial x \tag{30}$$

Eq. (26) acquires the form

$$\frac{v_i P'_{L,R}}{kT_0} dR + \frac{v_i P'_{L,x} + \mu'_i}{kT_0} dx - d \ln P_{0i} = 0 \tag{31}$$

We have the vector field $\mathbf{F} = (v_i P'_{L,R} / kT_0)\mathbf{i} + ((v_i P'_{L,x} + \mu'_i)/kT_0)\mathbf{j} - \mathbf{k}$ in the space $\mathbf{r} = (R, x, \ln P_{0i})$ and

$$rot\mathbf{F} = \frac{1}{kT_0} \left( -\frac{\partial(v_i P'_{L,x} + \mu'_i)}{\partial \ln P_{0i}}, \frac{\partial(v_i P'_{L,R})}{\partial \ln P_{0i}}, \frac{\partial(v_i P'_{L,x} + \mu'_i)}{\partial R} - \frac{\partial(v_i P'_{L,R})}{\partial x} \right) \tag{32}$$

If the dependence of the surface tension on $P_0$ is neglected, then the first and second components of $rot\mathbf{F}$ are equal to zero; the third component can be transformed as follows:

$$(rot\mathbf{F})_3 = -(\mathbf{F}rot\mathbf{F}) = \frac{1}{kT_0}\left(\frac{\partial(\upsilon_i P'_{L,x} + \mu'_i)}{\partial R} - \frac{\partial(\upsilon_i P'_{L,R})}{\partial x}\right) = \frac{\upsilon_i}{kT_0}\frac{\partial^2 P_L}{\partial x \partial R} \neq 0 \qquad (33)$$

So, Eq. (31) is not integrated by one relation. It is integrated by two relations $x(R)$ and $z(R)$, where $z \equiv \ln P_{0i}$. The function $x(R)$ id defined arbitrarily, Eq. (31) then converts to an ordinary differential equation for $z(R)$. In other words, we integrate Eq. (31) with respect to $R$ as an ordinary differential equation keeping in mind that the quantity $x$ is a function of $R$. The dependence $x(R)$ in fact is not arbitrary; this is the dependence of the critical nucleus composition on its radius which has to be determined from the final generalized Kelvin's equations. When $R$ changes from $\infty$ to the current $R$, $x$ changes from $x_\infty \equiv x(R=\infty)$ to $x(R)$. Integrating Eq. (31), we get

$$kT_0 \ln\frac{P_{0i}}{P_{0i}(x_\infty)} = \int_\infty^R \left\{\upsilon_i(x(R))\left[P'_{L,R} + P'_{L,x}\frac{dx}{dR}\right] + \mu'_i\frac{dx}{dR}\right\}dR = \int_0^{P_L} \upsilon_i(x(P_L))dP_L + \int_{x_\infty}^{x(R)} \mu'_i dx \qquad (34)$$

Consider the case, when $\upsilon_i$ *does not depend on* $x$ and the compositional dependence of the chemical potentials does not involve pressure (this means that $\mu'_i \equiv (\partial\mu_i/\partial x)_{T,P}$ does not involve $P_L$ as a parameter and thereby does not depend on $x$ via $P_L(x)$), i.e. the activities do not depend on pressure: $\mu_i(T_0,P,x) = \mu_i(T_0,P) + kT_0 \ln A_i(T_0,x)$. Then

$$\int_{x_\infty}^{x(R)} \mu'_i dx = \mu_i(x(R)) - \mu_i(x_\infty) = \Delta\mu_i(x(R)) - \Delta\mu_i(x_\infty) \qquad (35)$$

Employing $P_{0i}(x_\infty) = P_{si}(T_0)\exp(\Delta\mu_i(x_\infty)/kT_0)$, we get finally

$$P_{0i} = P_{si}(T_0)\exp\left\{\frac{2\upsilon_i\sigma(x)/R + \Delta\mu_i(T_0,x)}{kT_0}\right\} = P_{si}(T_0)A_i(T_0,x)\exp\left(\frac{2\upsilon_i\sigma(x)}{kT_0 R}\right) \qquad (36)$$

i.e. equation of the same form as Eq. (29). Whereas $\upsilon_i$ may depend on $x$ in Eq. (29), but $\sigma$ does not depend on it, in Eq. (36) the situation is reverse.

So, we have the system of equations

$$\begin{cases} x_0 P_0 = P_{s1}(T_0)A_1(T_0,x)\exp\left(\dfrac{2\upsilon_1\sigma(x)}{kT_0 R}\right) \\ (1-x_0)P_0 = P_{s2}(T_0)A_2(T_0,x)\exp\left(\dfrac{2\upsilon_2\sigma(x)}{kT_0 R}\right) \end{cases} \qquad (37)$$

Dividing the second equation by the first one, restoring asterisk and introducing the designations

$$\gamma_s \equiv \frac{P_{s2}}{P_{s1}},\ A_{0i} = \frac{P_{0i}}{P_{si}},\ A_0 \equiv \frac{A_{02}}{A_{01}},\ A \equiv \frac{A_2}{A_1},\ \upsilon_{21} \equiv \upsilon_2 - \upsilon_1,\ \frac{1-x_0}{x_0} = A_0\gamma_s$$

we get the following explicit dependence $R_*(x_*)$:

$$R_*(x_*)_{T_0,x_0} = \frac{2\upsilon_{21}\sigma(x_*)}{kT_0}\left(\ln\frac{A_0}{A(x_*,T_0)}\right)^{-1} \qquad (38)$$

from where the desired dependence $x_*(R_*)_{T_0,x_0}$ is obtained by inversion.

In a single-component case,

$$R_* = \frac{2\upsilon\sigma}{kT_0}\left(\ln\frac{P_0}{P_s}\right)^{-1} \tag{39}$$

The equality $A_0 = A(x_\infty, T_0)$ corresponding to $R_* = \infty$ determines the quantity $x_\infty$. The solution of Eq. (38) exists for $A_0 > A(x_*, T_0)$; thus at constant $T_0$ and $x_0$, the ratio $A_0 / A$ plays the role of "supersaturation" for binary nucleation in the sense of the single component supersaturation $P_0 / P_s$. Different pairs $(R_*, x_*)$ in Eq. (38) correspond to different values of the total pressure $P_0$ (which is determined from any of Eqs. (37) for the given $(R_*, x_*)$-values), just as $R_*$ in Eq. (39) depends on $P_0$.

If the dependence $x_*(R_*)_{T_0,P_0}$ (at constant $T_0$ and $P_0$) is required, then Eqs. (37) are added; the resulting equation of the form $P_0 = f(T_0, x_*, R_*)$ is a transcendental equation for the desired dependence. Different pairs $(R_*, x_*)$ in this case correspond to different values of $x_0$.

Returning to general Eq. (34) and doing in the similar way, we arrive at some *integral* equation for the dependence $x_*(R_*)_{T_0,x_0}$. Thus, some explicit and simple relations can be obtained only in the case when either $\sigma$ or $\upsilon_i$ does not depend on composition.

### 2.2.3. Application to ideal and regular solutions

In the present report, the approximation with *constant* $\upsilon_i$ *and* $\sigma$ is employed as the first step in constructing a self-consistent multivariable theory. This is a good approximation for ideal solutions; some estimates with a composition-dependent surface tension will be made for a regular solution for comparison. As an example of ideal solution, the *o*-xylene-*m*-xylene system is used; its parameters are given in Table 1. As an example of regular solution, the mixture of water with some hypothetical substance is used; the sulfuric acid parameters are employed for this substance (Table 2). This does not mean the attempt to describe adequately the properties of the $H_2O - H_2SO_4$ solution by the regular solution model; the aim is only to obtain a qualitative picture of the nucleation process. The two main properties of sulfuric acid – very small saturation pressure $P_{s2}$ and large heat of solution – result in some interesting peculiarities of the nucleation process discussed below.

We have for an ideal solution $A_1 = x$ and $A_2 = 1 - x$. Division of one of Eqs. (37) (with constant $\sigma$) by another gives

$$\frac{1 - x_0}{x_0} = \gamma_{R_*}(T_0)\frac{1 - x_*}{x_*} \tag{40}$$

from where the following relations between the compositions of the critical droplet and vapor are obtained:

$$x_*(x_0)_{T_0,R_*} = \frac{\gamma_{R_*} x_0}{1+(\gamma_{R_*}-1)x_0}\ ,\quad x_0(x_*)_{T_0,R_*} = \frac{x_*}{x_* + \gamma_{R_*}(1-x_*)}\ ,\quad \gamma_{R_*}(T_0) \equiv \gamma_s(T_0)\,\mathrm{e}^{\frac{2\upsilon_{21}\sigma}{kT_0R_*}} \tag{41}$$

With the help of equation $(1-x_0)/x_0 = A_0\gamma_s$, we get also

$$x_*(R_*)_{T_0,x_0} = \left[1 + A_0\,\mathrm{e}^{-\frac{2\upsilon_{21}\sigma}{kT_0R_*}}\right]^{-1},\quad R_*(x_*)_{T_0,x_0} = \frac{2\upsilon_{21}\sigma}{kT_0}\left[\ln A_0 \frac{x_*}{1-x_*}\right]^{-1} \tag{42}$$

Adding Eqs. (37), one obtains

$$x_*(R_*)_{T_0,P_0} = \frac{P_0 - P_{s2}\exp(2\upsilon_2\sigma/kT_0R_*)}{P_{s1}\exp(2\upsilon_1\sigma/kT_0R_*) - P_{s2}\exp(2\upsilon_2\sigma/kT_0R_*)} \tag{43}$$

Finally, the following equation for determining the critical radius $R_*(T_0,P_0,x_0)$ corresponding to the given vapor state can be derived from the above relations:

$$\mathrm{e}^{\frac{2\upsilon_2\sigma}{kT_0R_*}} = \frac{P_0}{P_{s2}(T_0)}\left[1 + \left(\gamma_{R_*}(T_0) - 1\right)x_0\right] \tag{44}$$

According to the regular solution definition, its excessive volume and entropy (the differences of the values of these quantities for regular and ideal solutions) are equal to zero; the entropy itself is the same as that for an ideal solution, i.e.

$$\Delta\upsilon = 0\,,\quad \Delta s = 0\,,\quad s(T,P,x) = \tilde{s}_1(T,P)x + \tilde{s}_2(T,P)(1-x) - k\left[x\ln x + (1-x)\ln(1-x)\right] \tag{45}$$

where $\tilde{s}_i(T,P)$ is the entropy of pure liquid $i$.

The distinction from ideal solution is only due to the heat of solution, so that the excessive enthalpy and Gibbs' potential (per one molecule) are

$$\Delta h = \Delta g = \Omega x(1-x) \tag{46}$$

where $\Omega$ is the characteristic parameter.

It is clear that the parameter $\Omega$ must be constant in order to satisfy the regular solution definition. If the dependence $\Omega(T,P)$ is accepted, then the finite values of $\Delta\upsilon$ and $\Delta s$ are inevitably appear:

$$\Delta\upsilon = \left(\frac{\partial\Delta g}{\partial P}\right)_{T,x} = x(1-x)\left(\frac{\partial\Omega}{\partial P}\right)_T\,,\quad \Delta s = -\left(\frac{\partial\Delta g}{\partial T}\right)_{P,x} = -x(1-x)\left(\frac{\partial\Omega}{\partial T}\right)_P\,,\quad \Delta h = x(1-x)\left[\Omega - T\left(\frac{\partial\Omega}{\partial T}\right)_P\right] \tag{47}$$

This version can be considered as some generalization of the regular solution concept and a more realistic model. However, here the standard definition with constant $\Omega$ is employed.

So, the chemical potentials and activities are

$$\mu_1(T_0,P,x) = \tilde{\mu}_1(T_0,P) + kT_0\left[\ln x + \omega(1-x)^2\right],\quad \mu_2(T_0,P,x) = \tilde{\mu}_2(T_0,P) + kT_0\left[\ln(1-x) + \omega x^2\right] \tag{48a}$$

$$A_1(x) = x\,\mathrm{e}^{\omega(1-x)^2}\,,\quad A_2(x) = (1-x)\,\mathrm{e}^{\omega x^2}\,,\quad \omega \equiv \Omega/kT_0 \tag{48b}$$

The case of $\omega < 0$ is considered here (the heat of solution is released); Eq. (38) acquires the form

$$R_*(x_*)_{T_0,x_0} = \frac{2\upsilon_{21}\sigma}{kT_0}\left(\ln\frac{A_0}{A(x_*)}\right)^{-1},\quad A(x) = \frac{1-x}{x}\,\mathrm{e}^{|\omega|(1-2x)} \tag{49}$$

The $P_0$-value corresponding to the pair $(x_*, R_*)$ is obtained from any of Eqs. (37), say, the first:

$$P_0 = \frac{1}{x_0} P_{s1} \, e^{\frac{2v_1\sigma}{kT_0 R_*}} \, x_* \, e^{-|\omega|(1-x_*)^2} \tag{50}$$

Eq. (41) is replaced by the following one:

$$x_0(x_*)_{T_0,R_*} = \frac{x_*}{x_* + \gamma_{R_*}(1-x_*) e^{|\omega|(1-2x_*)}} \tag{51}$$

The parameter $\omega$ for the $H_2O - H_2SO_4$ system evaluated form the data on the heat of solution [30] according to Eq. (46) depends on $x$, which means that this system is not described by the regular solution model; $\omega = 25$ as some characteristic value is employed below for calculating the nucleation kinetics .

The dependence $x_*(R_*)_{T_0,x_0}$ according to Eq. (49) with $\sigma = \sigma(0.586)$ is shown in Fig. 1b for $A_0 = 0.1$ and the corresponding partial pressures given in Table 2. The function $A(x)$ is descending, hence, the dependence $R_*(x_*)_{T_0,x_0}$ exists in the interval $x_\infty < x_* < 1$. The saturation pressure of sulfuric acid is very small, so that $\gamma_s \sim 10^{-5}$. From the equality $P_{02} / P_{01} = A_0 \gamma_s$ it is seen that extremely small values of the acid partial pressure $P_{02}$ are obtained for sufficiently small values of $A_0$ (e.g. $P_{02} / P_{01} \sim 10^{-8}$ for $A_0 = 10^{-3}$), nevertheless, nucleation occurs even for such small acid concentrations – Eq. (49) gives the sufficiently small $R_*$-values required for nucleation. The physical reason of this phenomenon is a negative heat of solution which means that the free energy of the system decreases, when the solution is formed. In unary nucleation, the system free energy decreases when the postcritical droplet grows; $R_* \to 0$ only at the highest degree of supersaturation (near the spinodal). In the considered case, even the formation of the simplest complex consisting of two molecules of different species already reduces the free energy – Eq. (49) formally gives non-physically small values of $R_*$ at usual atmospheric conditions, when each of the species is undersaturated in the sense of a pure substance.

The compositional dependence of the solution surface tension $\sigma(x)$ was obtained by the cubic spline interpolation of nine experimental points at $T_0 = 20\ ^\circ C$ [30]; it is not monotonic and has a maximum, $\sigma_{\max} = \sigma(0.743) = 77.29\ erg / cm^2$ (Fig. 1a). The dependence $x_*(R_*)_{T_0,x_0}$ calculated according to Eq. (38) is shown in Fig.1b by dashed line.

## 3. Kinetics of binary nucleation

### 3.1. Droplet motion equations in the $(V, x, T)$-space

The general form of the motion equations in the vicinity of the saddle point is

$$\begin{cases} \dot{V} = -z_{VV}(V - V_*) - z_{Vx}(x - x_*) - z_{VT}(T - T_0) \\ \dot{x} = -z_{xV}(V - V_*) - z_{xx}(x - x_*) - z_{xT}(T - T_0) \\ \dot{T} = -z_{TV}(V - V_*) - z_{Tx}(x - x_*) - z_{TT}(T - T_0) \end{cases} \quad (52)$$

Equations for the stable variables can be also represented as [25]

$$\begin{cases} \dot{x} = a_x\dot{V} - \lambda_{xx}(x - x_*) - \lambda_{xT}(T - T_0) \\ \dot{T} = a_T\dot{V} - \lambda_{Tx}(x - x_*) - \lambda_{TT}(T - T_0) \end{cases} \quad (53)$$

Such a representation allows identifying the characteristic parameters $\lambda_{ik}$ governing the nucleation kinetics; in particular, as shown below, the kinetic limits are determined just by these parameters.

From comparison, the matrix $\mathbf{Z}$ is

$$\mathbf{Z} = \begin{pmatrix} z_{VV} & z_{Vx} & z_{VT} \\ a_x z_{VV} & a_x z_{Vx} + \lambda_{xx} & a_x z_{VT} + \lambda_{xT} \\ a_T z_{VV} & a_T z_{Vx} + \lambda_{Tx} & a_T z_{VT} + \lambda_{TT} \end{pmatrix} \quad (54)$$

The conditions of symmetry of the matrix $\mathbf{D} = kT_0\,\mathbf{Z}\mathbf{H}^{-1}$ result in equations

$$z_{Vx}h_{xx}^{-1} = a_x z_{VV}h_{VV}^{-1}, \quad z_{VT}h_{TT}^{-1} = a_T z_{VV}h_{VV}^{-1}, \quad h_{TT}^{-1}(a_x z_{VT} + \lambda_{xT}) = h_{xx}^{-1}(a_T z_{Vx} + \lambda_{Tx}) \quad (55)$$

from where

$$a_x = \frac{z_{Vx}}{z_{VV}}\frac{h_{VV}}{h_{xx}}, \quad a_T = \frac{z_{VT}}{z_{VV}}\frac{h_{VV}}{h_{TT}} \quad (56)$$

The third of Eqs. (55) with account for Eqs. (56) converts to Onsager's relation for the matrix $\Lambda$: $\lambda_{xT}h_{TT}^{-1} = \lambda_{Tx}h_{xx}^{-1}$.

Eq. (53) for $\dot{T}$ can be also transformed as follows:

$$\dot{T} = a_T\dot{V} + b_T(\dot{x})_V + \lambda_{EE}(T - T_0), \quad (\dot{x})_V = -\lambda_{xx}(x - x_*) - \lambda_{xT}(T - T_0) \quad (57)$$

from where

$$\lambda_{Tx} = b_T\lambda_{xx}, \quad \lambda_{TT} = b_T\lambda_{xT} + \lambda_{EE} = \frac{\lambda_{Tx}\lambda_{xT}}{\lambda_{xx}} + \lambda_{EE}; \quad \lambda_{EE} = \frac{\det\Lambda}{\lambda_{xx}}, \quad \det\Lambda = \lambda_{xx}\lambda_{TT} - \lambda_{xT}\lambda_{Tx} = \lambda_{xx}\lambda_{EE} \quad (58)$$

### 3.2. Derivation of the matrix $\mathbf{Z}$

Eqs. (4) are used for deriving the matrix $\mathbf{Z}$. The values of equilibrium pressures $P_{ei}$ are given by Eq. (29) or Eq. (36) for arbitrary $R$ and $T$:

$$P_{ei}(R, x, T) = c_i\,\mathrm{e}^{\frac{1}{kT}\left\{-q_i + \frac{2\upsilon_i\sigma}{R} + \Delta\mu_i(R,x,T)\right\}} \quad (59)$$

$P_{0i}$ in Eqs. (4) are the current partial pressures in the vapor. Also the following equations are employed:

$$\dot{V} = \upsilon_1\dot{N}_1 + \upsilon_2\dot{N}_2 \quad (60a)$$

$$\dot{x} = \frac{1 - x}{N}\dot{N}_1 - \frac{x}{N}\dot{N}_2 \quad (60b)$$

Expanding the right-hand sides of Eqs. (4) near the saddle point up to linear terms (the quantities $\sigma$ and $\upsilon$ are assumed constant) and employing Eq. (60a), we find the elements $z_{VV}$, $z_{Vx}$, and $z_{VT}$. After simple transformations, they can be represented as follows:

$$z_{VV} = -(\xi_1 + \xi_2) P_L^* , \quad \xi_1 \equiv \frac{\beta_1 u_1 x_0}{\zeta_1^2 P_0 R_*} , \quad \xi_2 \equiv \frac{\beta_2 u_2 (1 - x_0)}{\zeta_2^2 P_0 R_*} , \quad \zeta_i \equiv \frac{\upsilon_0}{\upsilon_i} \tag{61a}$$

$$z_{Vx} = \frac{3 V_* \mu'_{1*}}{1 - x_*} \left[ (1 - x_*) \frac{\xi_1}{\upsilon_1} - x_* \frac{\xi_2}{\upsilon_2} \right] \tag{61b}$$

$$z_{VT} = 3kV_* \left[ \frac{\xi_1}{\upsilon_1} \bar{q}_1 + \frac{\xi_2}{\upsilon_2} \bar{q}_2 \right] , \quad \bar{q}_i \equiv \frac{q_i - kT_0/2 - 2\upsilon_i \sigma / R_* - \Delta h_i}{kT_0} \tag{61c}$$

where $\Delta h_i = \Delta \mu_i + T_0 \Delta s_i$ is the partial (differential) heat of mixing; $\Delta s_i \equiv [s_i(x) - \tilde{s}_i] = -(\partial \Delta \mu / \partial T)_{P,x}$. For both ideal and regular solutions,

$$s_i(P,T,x) = \tilde{s}_i(P,T) - k \ln x_i , \quad s_i(x) - \tilde{s}_i = -k \ln x_i \tag{62}$$

Hence,

$$\frac{\Delta h_i}{kT_0} = \begin{cases} 0 , & \text{ideal} \\ \omega (1 - x_i)^2 , & \text{regular} \end{cases} \tag{63}$$

The integral heat of mixing is $\Delta h = x \Delta h_1 + (1 - x) \Delta h_2$; $\Delta h = 0$ for ideal solution, as it must, and $\Delta h = \Omega x (1 - x)$ for regular solution, in accordance with Eq. (46). So, the quantity $\bar{q}_i$ for ideal solution has the same form as in a single component case. For regular solution, $-\Delta h > 0$, since $\Omega < 0$, i.e. this addend increases $\bar{q}_i$. As was shown in unary nucleation [17], the quantity $\bar{q}$ determines the heating of the droplet in the process of its growth due to the release of the condensation heat $q$. Here this effect is enhanced by the additional release of the mixing heat $\Delta h_i$.

In the case of a composition-dependent surface tension, Eq. (61b) would include the terms $l_i \equiv 2\upsilon_i (d\sigma / dx)_* / kT_0 R_*$ alongside with $\mu'_{i*}$. The assessment for the interpolated dependence $\sigma(x)$ mentioned above and Table 2 parameters gives $\mu'_{1*} / kT_0 \approx 22.4$ and $-\mu'_{2*} / kT_0 \approx 31.7$, whereas $l_1 \approx 1.3$ and $l_2 \approx 3.8$.

The elements of the second row of the matrix $\mathbf{Z}$ are obtained from Eq. (60b) in the similar way. In particular, the element $z_{xV}$ is

$$z_{xV} = -\frac{P_L^*}{N_*} \left[ (1 - x_*) \frac{\xi_1}{\upsilon_1} - x_* \frac{\xi_2}{\upsilon_2} \right] \tag{64}$$

On the other hand, the value of $z_{xV}$ is dictated by the symmetry conditions; Eqs. (54) and (56) give $z_{xV} = a_x z_{VV} = z_{Vx} h_{VV} / h_{xx}$. Substituting here $h_{VV}$ and $h_{xx}$ from Eq.(24) and employing Eq. (61b), we arrive at the same Eq. (64), i.e. the necessary condition of *self-consistency* of the theory is satisfied. From the known $z_{xV}$, an explicit form of the quantity $a_x = z_{xV} / z_{VV}$ is found. Similarly, equation $z_{TV} = z_{VT} h_{VV} / h_{TT}$ determines $z_{TV}$ and then $a_T = z_{TV} / z_{VV}$:

$$a_x = \frac{(1-x_*)\xi_1 / \upsilon_1 - x_* \xi_2 / \upsilon_2}{N_*(\xi_1 + \xi_2)}, \quad a_T = \frac{kT_0}{C_V^*} \frac{\bar{q}_1 \xi_1 / \upsilon_1 + \bar{q}_2 \xi_2 / \upsilon_2}{\xi_1 + \xi_2} \tag{65}$$

The matrix $\Lambda$ elements $\lambda_{xx}$ and $\lambda_{xT}$ are found from the known $z_{xx}$ and $z_{xT}$: $\lambda_{xx} = z_{xx} - a_x z_{Vx}$, $\lambda_{xT} = z_{xT} - a_T z_{VT}$. The element $\lambda_{Tx}$ is found from the symmetry condition $\lambda_{Tx} = \lambda_{xT} h_{xx} / h_{TT}$ and determines $z_{Tx} = a_T z_{Vx} + \lambda_{Tx}$; the parameter $\lambda_{TT}$ is calculated according to Eq. (58). The second condition of self-consistency is the physical meaning of the equation for $\dot{T}$ as an *energy balance equation*, from where the remaining parameter $\lambda_{EE}$ is determined. As a result, the matrices $\mathbf{Z}$ and $\Lambda$ are

$$\mathbf{Z} = \begin{pmatrix} -(\xi_1+\xi_2)P_L^* & \frac{3V_*\mu'_{1*}}{1-x_*}\left[(1-x_*)\frac{\xi_1}{\upsilon_1} - x_*\frac{\xi_2}{\upsilon_2}\right] & 3kV_*\left[\frac{\xi_1}{\upsilon_1}\bar{q}_1 + \frac{\xi_2}{\upsilon_2}\bar{q}_2\right] \\ -\frac{P_L^*}{N_*}\left[(1-x_*)\frac{\xi_1}{\upsilon_1} - x_*\frac{\xi_2}{\upsilon_2}\right] & \frac{3V_*\mu'_{1*}}{N_*(1-x_*)}\left[(1-x_*)^2\frac{\xi_1}{\upsilon_1^2} + x_*^2\frac{\xi_2}{\upsilon_2^2}\right] & \frac{3kV_*}{N_*}\left[(1-x_*)\frac{\xi_1}{\upsilon_1^2}\bar{q}_1 - x_*\frac{\xi_2}{\upsilon_2^2}\bar{q}_2\right] \\ -\frac{P_L^* kT_0}{C_V^*}\left[\frac{\xi_1}{\upsilon_1}\bar{q}_1 + \frac{\xi_2}{\upsilon_2}\bar{q}_2\right] & \frac{3V_*\mu'_{1*}}{(1-x_*)}\frac{kT_0}{C_V^*}\left[(1-x_*)\frac{\xi_1}{\upsilon_1^2}\bar{q}_1 - x_*\frac{\xi_2}{\upsilon_2^2}\bar{q}_2\right] & \frac{3k^2T_0V_*}{C_V^*}\left[\left(\frac{\bar{q}_1}{\upsilon_1}\right)^2\xi_1 + \left(\frac{\bar{q}_2}{\upsilon_2}\right)^2\xi_2\right] + \lambda_{EE} \end{pmatrix} \tag{66}$$

$$\Lambda = \begin{pmatrix} \frac{3V_*\mu'_{1*}}{N_*(1-x_*)}\frac{\xi_1\xi_2}{(\xi_1+\xi_2)}\frac{\upsilon_*^2}{\upsilon_1^2\upsilon_2^2} & \frac{3kV_*}{N_*}\frac{\xi_1\xi_2}{(\xi_1+\xi_2)}\left(\frac{\bar{q}_1}{\upsilon_1} - \frac{\bar{q}_2}{\upsilon_2}\right)\frac{\upsilon_*}{\upsilon_1\upsilon_2} \\ \frac{3V_*\mu'_{1*}kT_0}{(1-x_*)C_V^*}\frac{\xi_1\xi_2}{(\xi_1+\xi_2)}\left(\frac{\bar{q}_1}{\upsilon_1} - \frac{\bar{q}_2}{\upsilon_2}\right)\frac{\upsilon_*}{\upsilon_1\upsilon_2} & \frac{3k^2T_0V_*}{C_V^*}\frac{\xi_1\xi_2}{(\xi_1+\xi_2)}\left(\frac{\bar{q}_1}{\upsilon_1} - \frac{\bar{q}_2}{\upsilon_2}\right)^2 + \lambda_{EE} \end{pmatrix} \tag{67}$$

It is seen that always $\det\Lambda > 0$ due to Eq. (58) and $\lambda_{xx}, \lambda_{\mathrm{EE}} > 0$. Thus the condition of positiveness of $\det\Lambda$ does not impose any restrictions on nucleation kinetics.

Now all equations are available in order to represent Eq. (57) for $\dot{T}$ as an energy balance equation. From Eq. (58),

$$b_T = \frac{kT_0}{C_V^*}\frac{\upsilon_1\upsilon_2 N}{\upsilon}\left(\frac{\bar{q}_1}{\upsilon_1} - \frac{\bar{q}_2}{\upsilon_2}\right) \tag{68}$$

The condition of constant $V$ (or $\dot{V} = 0$), according to Eq. (60a), results in the equality $\upsilon_1\dot{N}_1 = -\upsilon_2\dot{N}_2$. Employing this equality in Eq. (60b), we have

$$(\dot{x})_V = \frac{\upsilon}{\upsilon_2 N}\dot{N}_1 = -\frac{\upsilon}{\upsilon_1 N}\dot{N}_2 \tag{69}$$

The product $b_T(\dot{x})_V$ can be represented in the "symmetric" form as follows:

$$b_T(\dot{x})_V = -\frac{kT_0}{C_V^*}\left(\frac{\bar{q}_1}{\upsilon_1} - \frac{\bar{q}_2}{\upsilon_2}\right)\frac{\left[-\upsilon_1\dot{N}_1\xi_2 + \upsilon_2\dot{N}_2\xi_1\right]}{\xi_1+\xi_2}\Bigg|_{\upsilon_1\dot{N}_1=-\upsilon_2\dot{N}_2} \tag{70}$$

Eq. (57) with account for Eqs. (60a), (65), and (70) converts to the following one:

$$C_V^*\dot{T} = kT_0(\bar{q}_1\dot{N}_1 + \bar{q}_2\dot{N}_2) - \lambda_{EE}C_V^*(T - T_0) \tag{71a}$$

or

$$dE = dQ + dW_{drop}\,,\;\; dE = C_V dT - \frac{kT_0}{2}dN\,,\;\; dW_{drop} = -kT_0 dN - \frac{2\upsilon_1\sigma}{R}dN_1 - \frac{2\upsilon_2\sigma}{R}dN_2\,,$$

$$dQ = (q_1 - \Delta h_1)dN_1 + (q_2 - \Delta h_2)dN_2 - \lambda_{TT}C_V(T - T_0)dt\,;\;\; -kT_0 = -P_0\upsilon_0 \approx P_0(\upsilon_i - \upsilon_0) \tag{71b}$$

Eq. (71a) is the expected energy balance equation; $\dot{E} = C_V^*\dot{T}$ is the critical droplet energy change per unit time, $kT_0(\bar{q}_1\dot{N}_1 + \bar{q}_2\dot{N}_2)$ is the energy change due to the release of the condensation and mixing heats, $-\lambda_{EE}C_V^*(T - T_0)$ is the energy change due to the heat exchange with the vapor according to Newton's law. Hence,

$$\lambda_{EE} = \frac{4\pi R_*^2}{C_V^*}\alpha = \frac{3\alpha}{c_V\rho_* R_*}\,,\;\; \rho_* \equiv \frac{1}{\upsilon_*}\,,\;\; c_V = \frac{C_V^*}{N_*} \tag{72}$$

where $\alpha$ is the heat transfer coefficient [20],

$$\alpha = \beta_{\varepsilon 1}(1-\beta_1)\rho_{01}u_1(c_{V1}^0 + k/2) + \beta_{\varepsilon 2}(1-\beta_2)\rho_{02}u_2(c_{V2}^0 + k/2) \tag{73}$$

Here $\beta_{\varepsilon i}$ are the thermal accommodation coefficients for each species; $\rho_{oi}$ are the partial densities of the vapors; $c_{Vi}^0$ are the heat capacities of the vapors. If a passive gas is present in the system, then the corresponding term is added to Eq. (73).

It should be noted that $\dot{N}_i$, according to Eq. (4), is the result of action of the two opposite processes – condensation and evaporation. Hence, $\dot{N}_i > 0$ and the droplet is *heated*, when it grows in both components. This fact should be kept in mind, when Eq. (71a) is analyzed. The last term in Eq. (71a) describes the relaxation of the droplet temperature to its equilibrium value due to the heat exchange with the vapor; this heat exchange is performed by reflected molecules, accordingly, Eq. (73) contains the fractions $(1-\beta_i)$ of such molecules. The droplet temperature plays a crucial role in its growth, which is seen from Eq. (59). We have for the *o*-xylene-*m*-xylene system $q_i/kT_0 \approx 15$, $2\upsilon_i\sigma/kT_0R_* \approx 2.2$, i.e. the first term dominates. When the droplet temperature increases, the pressure $P_{ei}$ increases also and $\dot{N}_i$ decreases. At some droplet temperature, its growth with respect to the $i$-th species is stopped and continues only after cooling due to interaction with the vapor. Thus, the droplet growth is controlled by its temperature to a great extent; this fact causes the *nonisothermal effect* in nucleation considered below.

The droplet growth (the increase of $R$) increases both $\dot{N}_1$ and $\dot{N}_2$. The change in the droplet composition also affects its growth velocity via $\Delta\mu_i(x)$, according to Eq. (59). When $x_i$ decreases, $P_{0i}$ decreases also and $\dot{N}_i$ increases. In other words, the deviation of the $i$-th species concentration from its equilibrium value $x_{i*}$ to smaller values causes the increase of the growth rate with respect to this species; and conversely, the deviation of $x_i$ to larger values causes the decrease of the corresponding growth rate. However, this effect works together with the above temperature effect, so that e.g. the increase of the growth rate in the first case can be diminished by the droplet heating.

Eq. (71b) is *the first law of thermodynamics* for the droplet. Thereby the relation between Onsager's symmetry conditions (Eqs. (55)) and the first law of thermodynamics revealed earlier in bubble nucleation [25] is also confirmed in droplet nucleation. Eq. (71b) for $dQ$ is the heat balance for the droplet: the full heat received by the droplet consists of the condensation and the mixing heats minus the heat given to the vapor ($T > T_0$). The origin of $(-kT_0)$ in $dW_{drop}$ and $(-kT_0/2)$ in $dE$ resulting in the quantity $(-kT_0/2)$ in Eq. (61c) for $\overline{q}_i$ is explained in Ref. [16] as the work done to keep the vessel at constant pressure and the energy per one molecule which must be given back from the droplet to the vapor in order to keep the latter at constant temperature, respectively. Both these terms and, as a result, the quantity $(-kT_0/2)$ in $\overline{q}_i$ are written therein from physical reasoning. It is worthy to note that here Eq. (61c) for $\overline{q}_i$ is a consequence of Eqs. (4) and (59). The basic model of nucleus formation developed in Ref. [25] and resulting in Eqs. (8) and (9) implies that both the temperature $T_0$ and the pressure $P_0$ of the parent phase remain *constant* at the nucleus formation and evolution. We see that the mentioned terms $(-kT_0)$ and $(-kT_0/2)$ in Eqs. (71b) ensure the fulfillment of this condition, i.e. there is the "feedback" between the basic model and resulting equations. All these facts also speak in favor of the self-consistency of the theory.

### 3.3. Reformulation of the theory in the $(N_1, N_2, E)$-variables

The aim of this Section is to show the consistency of the theory with the usually used approach in the $(N_1, N_2, E)$-variables. The matrix elements $z_{N_iN_j}$, $z_{N_iE}$, etc. are denoted below by $z_{ij}$, $z_{iE}$, etc. for brevity. So, the equations of motion have the form

$$\begin{cases} \dot{N}_1 = -z_{11}(N_1 - N_1^*) - z_{12}(N_2 - N_2^*) - z_{1E}(E - E_*) \\ \dot{N}_2 = -z_{21}(N_1 - N_1^*) - z_{22}(N_2 - N_2^*) - z_{2E}(E - E_*) \\ \dot{E} = -z_{E1}(N_1 - N_1^*) - z_{E2}(N_2 - N_2^*) - z_{EE}(E - E_*) \end{cases} \quad (74)$$

$$\dot{E} = a_{E1}\dot{N}_1 + a_{E2}\dot{N}_2 - \lambda_{EE}(E - E_*) \quad (75)$$

From these equations, one follows

$$z_{E1} = z_{11}a_{E1} + z_{21}a_{E2}, \quad z_{E2} = z_{12}a_{E1} + z_{22}a_{E2}, \quad z_{EE} = z_{1E}a_{E1} + z_{2E}a_{E2} + \lambda_{EE} \quad (76)$$

In order to derive the matrix $\mathbf{H}_{(N_1,N_2,E)}$, we have to express the derivatives of the function $W(N_1,N_2,E)$ via the derivatives of the function $W(V,x,T)$ with the help of relations $V(N_1,N_2)=\upsilon_1 N_1+\upsilon_2 N_2$, $x(N_1,N_2)=N_1/(N_1+N_2)$, and $dE=C_V^* dT$, e.g.

$$\frac{\partial W}{\partial N_1}=\frac{\partial W}{\partial V}\frac{\partial V}{\partial N_1}+\frac{\partial W}{\partial x}\frac{\partial x}{\partial N_1}=\upsilon_1\frac{\partial W}{\partial V}+\frac{N_2}{(N_1+N_2)^2}\frac{\partial W}{\partial x}$$

Computing the second derivative and taking it at the saddle point, we obtain the element $h_{11}=(\partial^2 W/\partial N_1^2)_*$. Doing in the similar way, we get

$$\mathbf{H}_{(N_1,N_2,E)}=\begin{pmatrix} \upsilon_1^2 h_{VV}+\frac{(1-x_*)^2}{N_*^2}h_{xx} & \upsilon_1\upsilon_2 h_{VV}-\frac{x_*(1-x_*)}{N_*^2}h_{xx} & 0 \\ \upsilon_1\upsilon_2 h_{VV}-\frac{x_*(1-x_*)}{N_*^2}h_{xx} & \upsilon_2^2 h_{VV}+\frac{x_*^2}{N_*^2}h_{xx} & 0 \\ 0 & 0 & \frac{1}{C_V^* T_0} \end{pmatrix} \tag{77}$$

Thus, the matrix is not diagonal, which reflects the physical equivalence of the variables $N_1$ and $N_2$.

From Eqs. (60a) and (60b) one obtains

$$\dot N_1=\frac{x\dot V+\upsilon_2 N\dot x}{\upsilon},\quad \dot N_2=\frac{(1-x)\dot V-\upsilon_1 N\dot x}{\upsilon} \tag{78}$$

These equations are employed for computing the matrix $\mathbf{Z}_{(N_1,N_2,E)}$. Doing as before, we have

$$\frac{\partial \dot N_1}{\partial N_1}=\frac{\partial \dot N_1}{\partial V}\frac{\partial V}{\partial N_1}+\frac{\partial \dot N_1}{\partial x}\frac{\partial x}{\partial N_1}=\upsilon_1\frac{\partial \dot N_1}{\partial V}+\frac{1-x}{N}\frac{\partial \dot N_1}{\partial x}$$

Substituting here Eq. (78) for $\dot N_1$ and taking the given derivative at the saddle point, we get the element $z_{11}=(\partial \dot N_1/\partial N_1)_*$. The remaining elements are calculated similarly; as a result,

$$z_{11}=\frac{\upsilon_1 x_*}{\upsilon_*}z_{VV}+\frac{x_*(1-x_*)}{\upsilon_* N_*}z_{Vx}+\frac{\upsilon_1\upsilon_2 N_*}{\upsilon_*}z_{xV}+\frac{(1-x_*)\upsilon_2}{\upsilon_*}z_{xx},\quad z_{12}=\frac{\upsilon_2 x_*}{\upsilon_*}z_{VV}-\frac{x_*^2}{\upsilon_* N_*}z_{Vx}+\frac{\upsilon_2^2 N_*}{\upsilon_*}z_{xV}-\frac{\upsilon_2 x_*}{\upsilon_*}z_{xx}$$

$$z_{1E}=\frac{x_*}{\upsilon_* C_V^*}z_{VT}+\frac{\upsilon_2 N_*}{\upsilon_* C_V^*}z_{xT},\quad z_{21}=\frac{(1-x_*)\upsilon_1}{\upsilon_*}z_{VV}+\frac{(1-x_*)^2}{\upsilon_* N_*}z_{Vx}-\frac{\upsilon_1^2 N_*}{\upsilon_*}z_{xV}-\frac{(1-x_*)\upsilon_1}{\upsilon_*}z_{xx}$$

$$z_{22}=\frac{(1-x_*)\upsilon_2}{\upsilon_*}z_{VV}-\frac{x_*(1-x_*)}{\upsilon_* N_*}z_{Vx}-\frac{\upsilon_1\upsilon_2 N_*}{\upsilon_*}z_{xV}+\frac{x_*\upsilon_1}{\upsilon_*}z_{xx},\quad z_{2E}=\frac{1-x_*}{\upsilon_* C_V^*}z_{VT}-\frac{\upsilon_1 N_*}{\upsilon_* C_V^*}z_{xT},\quad z_{EE}=z_{TT} \tag{79}$$

The symmetry conditions of the matrix $\mathbf{D}_{(N_1,N_2,E)}=kT_0\mathbf{Z}_{(N_1,N_2,E)}\mathbf{H}^{-1}_{(N_1,N_2,E)}$,

$$\mathbf{D}_{(N_1,N_2,E)}=\begin{pmatrix} z_{11}h_{22}-z_{12}h_{12} & -z_{11}h_{12}+z_{12}h_{11} & z_{1E}\Delta_{12}/h_{EE} \\ z_{21}h_{22}-z_{22}h_{12} & -z_{21}h_{12}+z_{22}h_{11} & z_{2E}\Delta_{12}/h_{EE} \\ z_{E1}h_{22}-z_{E2}h_{12} & -z_{E1}h_{12}+z_{E2}h_{11} & z_{EE}\Delta_{12}/h_{EE} \end{pmatrix} \tag{80}$$

give three equations

$$-z_{11}h_{12}+z_{12}h_{11}=z_{21}h_{22}-z_{22}h_{12} \tag{81a}$$

$$\begin{cases} z_{E1}h_{22}-z_{E2}h_{12}=z_{1E}\dfrac{\Delta_{12}}{h_{EE}} \\ -z_{E1}h_{12}+z_{E2}h_{11}=z_{2E}\dfrac{\Delta_{12}}{h_{EE}} \end{cases}, \quad \Delta_{12}\equiv \det \mathbf{H}_{12}\equiv \begin{vmatrix} h_{11} & h_{12} \\ h_{12} & h_{22} \end{vmatrix} = h_{VV}h_{xx}\frac{\upsilon_*^2}{N_*^2} \tag{81b}$$

Eq. (81a) after some transformations gives the symmetry condition $h_{VV}z_{Vx}=h_{xx}z_{xV}$ for the matrix $\mathbf{Z}$. Eqs. (81b) with account for Eqs. (76) give the system of equations for determining $a_{E1}$ and $a_{E2}$:

$$\begin{cases} (h_{22}z_{11}-h_{12}z_{12})a_{E1}+(h_{22}z_{21}-h_{12}z_{22})a_{E2}=z_{1E}\Delta_{12}/h_{EE} \\ (-h_{12}z_{11}+h_{11}z_{12})a_{E1}+(-h_{12}z_{21}+h_{11}z_{22})a_{E2}=z_{2E}\Delta_{12}/h_{EE} \end{cases} \tag{82}$$

Solving it by Cramer's rule, one obtains

$$a_{E1}=D_1/D\,, \quad a_{E2}=D_2/D \tag{83a}$$

where $D$ is the determinant of this system and

$$D_1\equiv\frac{\Delta_{12}}{h_{EE}}\begin{vmatrix} z_{1E} & h_{22}z_{21}-h_{12}z_{22} \\ z_{2E} & -h_{12}z_{21}+h_{11}z_{22} \end{vmatrix}, \quad D_2\equiv\frac{\Delta_{12}}{h_{EE}}\begin{vmatrix} h_{22}z_{11}-h_{12}z_{12} & z_{1E} \\ -h_{12}z_{11}+h_{11}z_{12} & z_{2E} \end{vmatrix} \tag{83b}$$

It is not difficult to get

$$D=\det\mathbf{H}_{12}\det\mathbf{Z}_{12}\,, \quad \det\mathbf{Z}_{12}\equiv\begin{vmatrix} z_{11} & z_{12} \\ z_{21} & z_{22} \end{vmatrix}=\begin{vmatrix} z_{VV} & z_{Vx} \\ z_{xV} & z_{xx} \end{vmatrix}=-\frac{3V_*\mu'_{1*}P_L^*}{N_*(1-x_*)}\xi_1\xi_2\frac{\upsilon_*^2}{\upsilon_1^2\upsilon_2^2} \tag{83c}$$

Calculation of $a_{Ei}$ by these formulae with the use of Eq. (66) yields the quite expected result:

$$a_{Ei}=kT_0\bar{q}_i \tag{84}$$

From equation $z_{EE}=z_{TT}=a_T z_{VT}+\lambda_{TT}$ and Eq. (76) for $\lambda_{EE}$, one obtains

$$\lambda_{TT}=\lambda_{EE}+\frac{3k^2T_0V_*}{C_V^*}\frac{\xi_1\xi_2}{\xi_1+\xi_2}\left[\frac{\bar{q}_1}{\upsilon_1}-\frac{\bar{q}_2}{\upsilon_2}\right]^2 \tag{85}$$

in accordance with Eq. (67). So, Eq. (75) is the energy balance equation, as it must.

The explicit form of the diffusion matrix, Eq. (80) is following:

$$\mathbf{D}_{(N_1,N_2,E)}=d_{11}\begin{pmatrix} 1 & 0 & a_{E1} \\ 0 & \gamma_d & \gamma_d a_{E2} \\ a_{E1} & \gamma_d a_{E2} & b^2+a_{E1}^2+\gamma_d a_{E2}^2 \end{pmatrix}, \quad \gamma_d\equiv\frac{d_{22}}{d_{11}}, \quad d_{ii}=4\pi R_*^2\beta_i u_i\rho_{0i}=\frac{3V_*kT_0}{\upsilon_i^2}\xi_i \tag{86a}$$

$$b^2\equiv\left[\beta_{\varepsilon1}\frac{1-\beta_1}{\beta_1}\left(c_{V1}^0+\frac{k}{2}\right)+\beta_{\varepsilon2}\frac{1-\beta_2}{\beta_2}\gamma_d\left(c_{V2}^0+\frac{k}{2}\right)\right]kT_0^2 \tag{86b}$$

The general form of the matrix $\mathbf{D}_{(N_1,N_2,E)}$ is the same as obtained in Ref. [9] (Eq. (50b) therein) from the linked-fluxes analysis. Eqs. (86a) and (86b) generalize the corresponding equations of a single component nucleation [17].

Finally, the explicit form of the matrix $\mathbf{Z}_{(N_1,N_2,E)}$ can be easier obtained from Eqs. (77) and (86a). In particular, the first two equations of motion are

$$\dot{N}_1 = \frac{d_{11}}{kT_0}\left\{\left[\frac{P_L^*}{3V_*}\upsilon_1^2 - \frac{1-x_*}{N_*}\mu_{1*}'\right](N_1 - N_1^*) + \left[\frac{P_L^*}{3V_*}\upsilon_1\upsilon_2 + \frac{x_*}{N_*}\mu_{1*}'\right](N_2 - N_2^*) - \frac{k}{C_V^*}\bar{q}_1(E - E_*)\right\} \quad (87a)$$

$$\dot{N}_2 = \frac{d_{22}}{kT_0}\left\{\left[\frac{P_L^*}{3V_*}\upsilon_1\upsilon_2 + \frac{x_*}{N_*}\mu_{1*}'\right](N_1 - N_1^*) + \left[\frac{P_L^*}{3V_*}\upsilon_2^2 - \frac{x_*^2}{(1-x_*)N_*}\mu_{1*}'\right](N_2 - N_2^*) - \frac{k}{C_V^*}\bar{q}_2(E - E_*)\right\} \quad (87b)$$

### 3.4. Kinetic limits

The negative eigenvalue $\kappa_1$ of the matrix $\mathbf{Z}$ is a solution of equation

$$\kappa^3 - (Sp\mathbf{Z})\kappa^2 + B\kappa - \det\mathbf{Z} = 0 \quad (88)$$

where $Sp\mathbf{Z} = z_{VV} + a_x z_{Vx} + \lambda_{xx} + a_T z_{VT} + \lambda_{TT}$, $\det\mathbf{Z} = z_{VV}\det\Lambda$,

$B = \det\mathbf{Z}_{Vx} + \det\mathbf{Z}_{VT} + \det\mathbf{Z}_{xT} = z_{VV}(\lambda_{xx} + \lambda_{TT}) + \det\mathbf{Z}_{xT}$;

$$\mathbf{Z}_{Vx} = \begin{pmatrix} z_{VV} & z_{Vx} \\ a_x z_{VV} & a_x z_{Vx} + \lambda_{xx} \end{pmatrix}, \quad \mathbf{Z}_{VT} = \begin{pmatrix} z_{VV} & z_{VT} \\ a_T z_{VV} & a_T z_{VT} + \lambda_{TT} \end{pmatrix}, \quad \mathbf{Z}_{xT} = \begin{pmatrix} a_x z_{Vx} + \lambda_{xx} & a_x z_{VT} + \lambda_{xT} \\ a_T z_{Vx} + \lambda_{Tx} & a_T z_{VT} + \lambda_{TT} \end{pmatrix} \quad (89)$$

are the main minors of the matrix $\mathbf{Z}$.

We have

$$\kappa_1 = \frac{1}{3}\left\{-2\sqrt{(Sp\mathbf{Z})^2 - 3B}\cos\left(\frac{\phi}{3} - \frac{\pi}{3}\right) + Sp\mathbf{Z}\right\} \quad (90a)$$

$$\cos\phi = -\frac{U}{2\sqrt{-(Y/3)^3}}, \quad U = -\frac{2}{27}(Sp\mathbf{Z})^3 + \frac{BSp\mathbf{Z}}{3} - \det\mathbf{Z}, \quad Y = B - \frac{(Sp\mathbf{Z})^2}{3} \quad (90b)$$

The condensation coefficients $\beta_i$ are obviously functions of composition; they enter in the matrix $\mathbf{Z}$ at the critical composition, $\beta_i(x_*)$. In order to simplify the problem and obtain some characteristic dependences, we put $\beta_1 = \beta_2 \equiv \beta$ for ideal *o*-xylene-*m*-xylene system and consider $\beta$ as a free parameter of the theory. The dependence $\kappa_1(\beta)$ which is in essence the *dependence of the nucleation rate on* $\beta$ is shown for this system in Fig. 2. It is seen that this dependence is asymmetric and has a maximum at $\beta \approx 0.2$. For comparison, the similar dependence for the unary nucleation in water vapor is shown; it is more asymmetric and the maximum is located at $\beta \approx 0.1$.

In the considered $H_2O - H_2SO_4$ system, the amount of the second component in the vapor is very small, $\rho_{02} << \rho_{01}$, $\xi_2/\xi_1 \sim 10^{-5}$, so that $\beta_2$ practically does not affect the nucleation and the water condensation coefficient $\beta_1$ is a free parameter of the theory. The function $\kappa_1(\beta_1)$ for this system is also shown in Fig. 2.

So, the kinetics of binary nucleation is governed by three parameters: $z_{VV}$, $\lambda_{xx}$, and $\lambda_{TT}$ (or $\lambda_{EE}$). In comparison with unary nucleation, the new kinetic parameter $\lambda_{xx}$ appears. Similarly to the parameter $\lambda_{TT}$ which characterizes the rate of temperature relaxation at constant $V$ and $x = x_*$, the parameter $\lambda_{xx}$ characterizes the rate of composition relaxation at constant $V$ and $T = T_0$. The parameter $z_{VV}$ characterizes the rate of droplet growth; the parameter $\lambda_{EE}$ is the rate of heat exchange between the droplet and the vapor. Different kinetic limits in binary nucleation are due to different limiting ratios between the above parameters.

Isothermal limit is $\lambda_{TT} \to \infty$; this means physically that the parameter $\lambda_{TT}$ is much greater than the other parameters in $Sp\mathbf{Z}$ (this limit is realized, in particular, in the case of $\beta_i \to 0$). We can use the formal way [26] – divide Eq. (88) by $\lambda_{TT}$ and take the mentioned limit. As a result,

$$\frac{Sp\mathbf{Z}}{\lambda_{TT}} \to 1, \quad \frac{B}{\lambda_{TT}} \to z_{VV} + a_x z_{Vx} + \lambda_{xx} = Sp\mathbf{Z}_{Vx}, \quad \frac{\det \mathbf{Z}}{\lambda_{TT}} \to z_{VV}\lambda_{xx}$$

and Eq. (88) converts to

$$\kappa^2 - (Sp\mathbf{Z}_{Vx})\kappa + \det \mathbf{Z}_{Vx} = 0 \tag{91}$$

which is the required isothermal equation of binary nucleation. Its solution is

$$\kappa_{iso} = \frac{1}{2}\left\{Sp\mathbf{Z}_{Vx} - \sqrt{(Sp\mathbf{Z}_{Vx})^2 - 4\det \mathbf{Z}_{Vx}}\right\} \tag{92}$$

As in unary nucleation, nonisothermal effect is characterized by the ratio

$$\psi \equiv \frac{I}{I_{iso}} = \frac{\kappa_1}{\kappa_{iso}} \tag{93}$$

The dependences $\psi(\beta)$ and $\psi(\beta_1)$ for the considered systems are shown in Fig. 3.

Further limits within the isothermal case are determined by interrelations between $z_{VV}$ and $\lambda_{xx}$. The *unary nucleation* limit $\lambda_{xx} \to \infty$, or $\lambda_{xx} >> |z_{VV}|, a_x z_{Vx}$, leads to the conditions $(Sp\mathbf{Z}_{Vx})^2 >> |\det \mathbf{Z}_{Vx}|$ and $Sp\mathbf{Z}_{Vx} > 0$; Eq. (92) has the following asymptotics in this case:

$$\kappa_{iso} = \frac{\det \mathbf{Z}_{Vx}}{Sp\mathbf{Z}_{Vx}} = \frac{z_{VV}\lambda_{xx}}{z_{VV} + a_x z_{Vx} + \lambda_{xx}} = z_{VV} \tag{94}$$

which is the one-dimensional value. Indeed, a large value of $\lambda_{xx}$ means that any deviation of the droplet composition from $x_*$ rapidly relaxes, i.e. the droplet grows with *constant composition* $x_*$. In other words, we have the one-dimensional growth; the variable $x$ falls out from consideration, only the variable $V$ remains.

In the opposite limit $\lambda_{xx} \to 0$, or $\lambda_{xx} << |z_{VV}|, a_x z_{Vx}$, under the same condition $Sp\mathbf{Z}_{Vx} > 0$, Eq. (94) yields

$$\kappa_{iso} = \frac{\det \mathbf{Z}_{Vx}}{Sp\mathbf{Z}_{Vx}} = \frac{z_{VV}}{z_{VV} + a_x z_{Vx}}\lambda_{xx} \tag{95}$$

The unary limit $\lambda_{xx} \to \infty$ of the $(V, x, T)$-theory can be considered similarly to the above limit $\lambda_{TT} \to \infty$. Eq. (88) in this limit converts to the equation of the single-component nonisothermal theory

$$\kappa^2 - (Sp\mathbf{Z}_{VT})\kappa + \det \mathbf{Z}_{VT} = 0 \tag{96}$$

considered in detail in the previous work. The solution $\kappa_x(\beta)$ of this equation is shown in Fig. 2. It is close to the dependence $\kappa_1(\beta)$ of the $(V,x,T)$-theory despite the fact that $\lambda_{xx}$ is not much greater than the other terms in $Sp\mathbf{Z}$; it is seen from Table 1 that $\lambda_{xx}$ and $|z_{VV}|$ are of the same order of magnitude. This means that the characteristic times of volume change and composition change are the same and the composition has a chance to adjust to the change in volume. As a consequence, the process does not differ greatly from a unary-nucleation process. The discussed dependences differ at $\beta < 0.4$ and coincide at $\beta > 0.4$ . Nonisothermal effect increases with increasing $\beta$ ($\lambda_{TT}$ decreases) and this fact is seen to make the theory unary, as a consequence of the relation between thermal and compositional effects discussed above.

The asymptotic form of $\kappa_1$ in the limits $\lambda_{xx} \to 0$ or $\lambda_{EE} \to 0$ within the $(V,x,T)$-theory can be obtained from Eq. (90a). The expected result is $\kappa_1 \sim \lambda_{xx}$ or $\kappa_1 \sim \lambda_{EE}$, respectively, similarly to Eq. (95). This result agrees with the general rule that the nucleation rate is determined by the *slowest* kinetic process in the system. Employing this fact and keeping in Eq. (88) only the terms of the first order of smallness, $B\kappa$ and $\det\mathbf{Z}$, we immediately get

$$\kappa_1 = \frac{\det\mathbf{Z}}{B} = \frac{z_{VV}\lambda_{EE}}{B}\lambda_{xx} \quad \text{or} \quad \kappa_1 = \frac{z_{VV}\lambda_{xx}}{B}\lambda_{EE} \qquad (97)$$

where the condition $B > 0$ is implied. The limit $\lambda_{EE} \to 0$ is realized at $\beta_i \to 1$ for both species. Both Eqs. (95) and (97) express the *limiting effect of kinetic processes on nucleation*: the nucleation rate tends to zero together with the corresponding kinetic parameter.

As was mentioned above, $\xi_2 << \xi_1$ in the considered $H_2O - H_2SO_4$ system; hence, according to Eq. (67), $\lambda_{xx} \sim \xi_2$ and $\lambda_{xx}$ is much less than the other terms in $Sp\mathbf{Z}$ (see Table 2). Thus the limit $\lambda_{xx} \to 0$ holds, and both Eqs. (90a) and (97) give the same values, e.g. $\kappa_1(0.1) = -1.193$. The dependence $\kappa_1(\beta_1)$ for this system shown in Fig. 2 has a plateau form; the difference in the $\kappa_1$-values corresponding to different $\beta_1$-values is in the third decimal place. Thus it is practically constant in the almost whole region $0 < \beta_1 < 1$ and tends to zero at $\beta_1 \to 0$ and $\beta_1 \to 1$. This means that the nucleation rate in the given system practically does not depend on both the condensation coefficients, however, $\beta_1$ cannot be put equal to unity, as is usually done. At $\beta_1 \to 1$, the nucleation rate tends to zero together with $\kappa_1$ as a consequence of nonisothermal effect ($\lambda_{EE} \to 0$ in the second Eq. (97)).

## 4. Steady state distribution function of droplets

### 4.1. Direction of the flux of droplets

For studying the distribution function of droplets, we use the dimensionless variables $y_1 = (V - V_*)/V_*$ , $y_2 = x - x_*$, and $y_3 = (T - T_0)/T_0$. As before, the matrix elements $z_{y_i y_j}$, etc. are denoted by $z_{ij}$, etc. The matrices $\mathbf{Z}$ and $\mathbf{H}$ change as follows:

$$\mathbf{Z} = \begin{pmatrix} z_{VV} & z_{Vx}/V_* & z_{VT}T_0/V_* \\ z_{xV}V_* & z_{xx} & z_{xT}T_0 \\ z_{TV}V_*/T_0 & z_{Tx}/T_0 & z_{TT} \end{pmatrix}, \quad h_{11} = \frac{h_{VV}V_*^2}{kT_0}, \quad h_{22} = \frac{h_{xx}}{kT_0}, \quad h_{33} = \frac{h_{TT}T_0}{k} \tag{98}$$

The flux direction is the unit eigenvector $\mathbf{e} = (c_{11}, c_{21}, c_{31})$ of the matrix $\mathbf{Z}$ corresponding to $\kappa_1$, i.e. $\mathbf{Ze} = \kappa_1 \mathbf{e}$. In expanded form, this matrix equation is the system of three equations for the vector $\mathbf{e}$ components. The third equation is a consequence of the first two equations, since $\det \mathbf{Z} = 0$ is the equation for eigenvalues. The first two equations are

$$\begin{cases} z_{12}c_{21} + z_{13}c_{31} = -(z_{11} - \kappa_1)c_{11} \\ (z_{22} - \kappa_1)c_{21} + z_{23}c_{31} = -z_{21}c_{11} \end{cases} \tag{99}$$

Solving this system by Cramer's rule, we have

$$c_{21} = c_{11}\frac{D_{21}}{D_0}, \quad c_{31} = c_{11}\frac{D_{31}}{D_0}, \quad D_{21} \equiv \begin{vmatrix} -(z_{11} - \kappa_1) & z_{13} \\ -z_{21} & z_{23} \end{vmatrix}, \quad D_{31} \equiv \begin{vmatrix} z_{12} & -(z_{11} - \kappa_1) \\ z_{22} - \kappa_1 & -z_{21} \end{vmatrix}, \quad D_0 \equiv \begin{vmatrix} z_{12} & z_{13} \\ z_{22} - \kappa_1 & z_{23} \end{vmatrix} \tag{100}$$

The vector $\mathbf{e}$ is a unit vector, hence, equation $c_{11}^2 + c_{21}^2 + c_{31}^2 = 1$ gives

$$c_{11} = \left[1 + \left(\frac{D_{21}}{D_0}\right)^2 + \left(\frac{D_{31}}{D_0}\right)^2\right]^{-1/2} \tag{101}$$

The quantities $c_{i1} = \cos\theta_i$ are the direction cosines of the vector $\mathbf{e}$, thus Eqs. (100) and (101) determine the flux direction.

Consider the limiting case when $\lambda_{xx}$ is much smaller than the other terms in $Sp\mathbf{Z}$. As was noted above, $\kappa_1 \sim \lambda_{xx}$ in this case; hence, $(\lambda_{xx} - \kappa_1) \sim \lambda_{xx}$. Expanding Eqs. (100) and (101) in this small parameter to linear terms with account for Eq. (54) for the matrix elements, the following asymptotics can be obtained:

$$\frac{D_{21}}{D_0} = a(1 + b\lambda_{xx}), \quad \frac{D_{31}}{D_0} = c\lambda_{xx}, \quad c_{11} = \left(1 + a^2\right)^{-1/2}\left[1 - \frac{ba^2}{1 + a^2}\lambda_{xx}\right], \quad c_{21} = a\left(1 + a^2\right)^{-1/2}\left[1 + \frac{b}{1 + a^2}\lambda_{xx}\right],$$

$$c_{31} = c\left(1 + a^2\right)^{-1/2}\lambda_{xx} \tag{102}$$

where $a$, $b$, and $c$ are some constants. It is seen that $c_{31} = \cos\theta_3 \to 0$ at $\lambda_{xx} \to 0$, i.e. $\theta_3 \to \pi/2$ and the flux lies in the $(V, x)$-plane. In other words, the nucleation process is *isothermal* in this limit. Further, $c_{11}^{\lim} = \cos\theta_1^{\lim} = \left(1 + a^2\right)^{-1/2}$, $c_{21}^{\lim} = \sin\theta_1^{\lim} = a\left(1 + a^2\right)^{-1/2}$, and $tg\,\theta_1^{\lim} = a$ is the limiting flux direction at $\lambda_{xx} \to 0$.

The limiting direction can be also determined within the $(V,x)$-theory with the help of the matrix $\mathbf{Z}_{Vx}$, Eq. (89). The flux direction in a $2D$ theory is [9]

$$tg\theta = \frac{1}{2z_{12}}\left\{(z_{22}-z_{11}) - \sqrt{(z_{22}-z_{11})^2 + 4z_{12}z_{21}}\right\} \quad (103)$$

Expanding this expression in the small parameter $\lambda_{xx}$ and implying $a_x z_{Vx} > |z_{VV}|$, we get the following limit at $\lambda_{xx} \to 0$:

$$tg\theta^{\lim} = \frac{|z_{11}|}{z_{12}} \quad (104)$$

from where $a = |z_{11}|/z_{12}$. Eq. (104) can be also applied to calculating the limiting flux direction in the single component $(V,T)$-theory at $\lambda_{TT} \to 0$ $(\beta \to 1)$, which is the strongly nonisothermal case. This limiting direction calculated numerically in the previous work coincides with that given by Eq. (104).

In our case, the matrix $\mathbf{H}$, Eq. (24), has diagonal form; hence,

$$W = W_* + \frac{1}{2}(h_{11}y_1^2 + h_{22}y_2^2 + h_{33}y_3^2) \quad (105)$$

Expansion of the work in the direction $\mathbf{e}$ (with the use of the directional derivative $(\mathbf{e}\nabla)W(\mathbf{r})$) is [9]

$$W = W_* + \frac{1}{2}h'_{11}(\mathbf{e})\mathbf{r}^2 = W_* + \frac{1}{2}h'_{11}(\mathbf{e})(y_1^2 + y_2^2 + y_3^2)_{\mathbf{r}\parallel\mathbf{e}} \quad (106)$$

where the vector $\mathbf{r} = (y_1, y_2, y_3)$ is parallel to the vector $\mathbf{e}$; $h'_{11}(\mathbf{e}) = h_{11}c_{11}^2 + h_{22}c_{21}^2 + h_{33}c_{31}^2$ is the element of the matrix $\mathbf{H}' = \mathbf{C}^T\mathbf{H}\mathbf{C}$; $\mathbf{C}$ is the matrix of transition to the new system of coordinates with the vector $\mathbf{e}$ as the first basic vector, accordingly, the components $(c_{11}, c_{21}, c_{31})$ form the first column of the matrix $\mathbf{C}$.

The quantity $h'_{11} \equiv \kappa_b(\mathbf{e})$ is the *curvature of the nucleation barrier*, or the curvature of the normal section of the $3D$ saddle surface in the direction $\mathbf{e}$. Eq. (105) is the expansion of the work in arbitrary direction of the vector $\mathbf{r}$. Applying it to the direction $\mathbf{e}$ and comparing with Eq. (106), we get again

$$\kappa_b(\mathbf{e}) = \frac{h_{11}y_1^2 + h_{22}y_2^2 + h_{33}y_3^2}{y_1^2 + y_2^2 + y_3^2} = h_{11}\cos^2\theta_1 + h_{22}\cos^2\theta_2 + h_{33}\cos^2\theta_3 = h_{11}c_{11}^2 + h_{22}c_{21}^2 + h_{33}c_{31}^2 \quad (107)$$

which is known from the differential geometry *Euler's formula* for the normal section curvature.

So, expansion of a multivariable work in the direction $\mathbf{e}$ yields *one-dimensional* Eq. (106), where the quantity $h'_{11}(\mathbf{e})$, the nucleation barrier curvature given by Euler's formula, generalizes the usual second derivative of the one-dimensional theory.

### 4.2. Distribution functions

The multivariable steady state distribution function $f_s(\mathbf{r})$ is [9]:

$$f_s(\mathbf{r}) = \frac{1}{2} f_{eq}(\mathbf{r}) erfc\left(\varphi(\mathbf{r})\right), \quad \varphi(\mathbf{r}) \equiv -\frac{\mathbf{eHr}}{\sqrt{2|\kappa_b(\mathbf{e})|}}, \quad \mathbf{r} = (y_1, y_2, y_3) \tag{108}$$

With account for the above equations,

$$\varphi(\mathbf{r}) \equiv -\frac{(c_{11}h_{11})y_1 + (c_{21}h_{22})y_2 + (c_{31}h_{33})y_3}{\sqrt{2|\kappa_b(\mathbf{e})|}} \tag{109}$$

The equilibrium distribution function is

$$f_e(y_1, y_2, y_3) = \zeta\rho_0 N_* \frac{\sqrt{h_{22}h_{33}}}{2\pi} \mathrm{e}^{-W_*' + \frac{1}{2}\left(|h_{11}| y_1^2 - h_{22}y_2^2 - h_{33}y_3^2\right)} \tag{110}$$

where $W_*' \equiv 4\pi\sigma R_*^2 / 3kT_0$ is the dimensionless nucleation work; $\zeta \sim \zeta_i$ is the factor from the statistical theory [10]. The nucleation rate values for both the considered systems given in Tables were calculated with account for this factor. Eqs. (108)-(110) express the distribution of droplets via the saddle surface geometry and the flux direction.

The droplet distributions on stable variables in the saddle point vicinity $-\delta < y_1 < \delta$ are defined as

$$f_s(y_2, y_3) = \frac{\int_{-\delta}^{\delta} f_s(y_1, y_2, y_3) dy_1}{\int_{-\delta}^{+\delta} dy_1 \int_{-\infty}^{+\infty}\int_{-\infty}^{+\infty} f_s(y_1, y_2, y_3) dy_2 dy_3}, \quad f_e(y_2, y_3) = \frac{\sqrt{h_{22}h_{33}}}{2\pi} \mathrm{e}^{-\frac{1}{2}\left(h_{22}y_2^2 + h_{33}y_3^2\right)} \tag{111}$$

where $\delta$ is the characteristic half-width of the critical region in $y_1$. Hence, the temperature and composition distributions are

$$f_s(y_3) = \int_{-\infty}^{+\infty} dy_2 f_s(y_2, y_3), \quad f_s(y_2) = \int_{-\infty}^{+\infty} dy_3 f_s(y_2, y_3), \quad f_e(y_3) = \sqrt{\frac{h_{33}}{2\pi}} \mathrm{e}^{-\frac{1}{2}h_{33}y_3^2}, \quad f_e(y_2) = \sqrt{\frac{h_{22}}{2\pi}} \mathrm{e}^{-\frac{1}{2}h_{22}y_2^2} \tag{112}$$

The mean steady-state overheat $\langle \Delta T \rangle$ and *enrichment* $\langle \Delta x \rangle$ are

$$\langle \Delta T \rangle = \langle T \rangle - T_0 = T_0 \int_{-\infty}^{+\infty} y_3 f_s(y_3) dy_3, \quad \langle \Delta x \rangle = \langle x \rangle - x_* = \int_{-\infty}^{+\infty} y_2 f_s(y_2) dy_2 \tag{113}$$

The functions $f_s(y_3)$ and $f_s(y_2)$ together with the corresponding equilibrium functions are shown in Fig.4 for the *o*-xylene-*m*-xylene system. It is seen that the function $f_s(y_3)$ is shifted relatively the function $f_e(y_3)$ to higher temperatures, i.e. there is the mean overheat of droplets which is shown in Fig. 5 as a function of the condensation coefficient. The smaller shift of the function $f_s(y_2)$ relatively $f_e(y_2)$ to *lower* compositions is observed in Fig. 4, which corresponds to the enrichment of droplets by the second species in the steady state; the absolute value of this enrichment is shown in Fig. 6 as a function of $\beta$. Despite the fact that the absolute value of $\langle \Delta x \rangle$ is small (of the order of $10^{-3}$), the relative values $\langle \Delta T \rangle / T_0$ and $\langle \Delta x \rangle / x_*$ are of the same order of magnitude. Obviously, $\langle \Delta T \rangle$ and $\langle \Delta x \rangle$ depend on $\delta$; $\delta = 0.25$ was put for both the considered systems. Both the overheat and enrichment are greater for larger droplets. The calculated flux direction at $\beta = 0.1$ is as follows: $\theta_1 = 2.095^\circ$, $\theta_2 = 91.111^\circ$, and $\theta_3 = 88.224^\circ$. It is seen that the flux is in negative direction with respect to the $x$-axis ($\theta_2 > \pi/2$), from where $\langle \Delta x \rangle$ is negative.

It was noted in the previous work that the dependences $\langle \Delta T(\beta) \rangle$, $\psi(\beta)$, etc. are sharp for $\beta$-values less than about $0.1$ and smooth for $\beta > 0.1$. Here the similar behavior takes place for the *o*-xylene-*m*-xylene system, but the characteristic value of $\beta$ is about $0.2$. It is seen from Fig. 2 that both these $\beta$-values correspond to the maximum nucleation rate.

The enrichment by the second species is maximum at $\beta \to 0$ (the isothermal limit); it is due to the difference in the growth rates $\dot{N}_1$ and $\dot{N}_2$. Mathematically the origin of the enrichment within the $2D$ $(V,x)$-theory operating with the matrix $\mathbf{Z}_{Vx}$, Eq. (89), is quite similar to the origin of the mean droplet overheat in the single component $(V,T)$-theory operating with the matrix $\mathbf{Z}_{VT}$.

As noted above, the droplet temperature and composition simultaneously affect the growth rates $\dot{N}_i$; from here the correlation of these quantities in the steady state follows which can be called the *kinetic correlation* in contrast to the thermodynamic correlation which is absent here ($h_{xT} = 0$ in Eq. (24)). The correlator $\langle (y_2 - \langle y_2 \rangle)(y_3 - \langle y_3 \rangle) \rangle$ can be calculated with the steady state distribution function $f_s(y_2, y_3)$; e.g. it is equal to $5.273 \times 10^{-6}$ for $\beta = 0.1$. Equation $(\dot{T})_V = b_T (\dot{x})_V + \lambda_{EE}(T - T_0)$ following from Eq. (57) and relating the changes in temperature and composition is the analytical expression of the kinetic correlation. This equation also leads to appearing the new parameter $\lambda_{EE}$ in addition to $\lambda_{TT}$. It is seen from Eq. (85) that the difference between these parameters is due to the differences in $q_i$ and in $\upsilon_i$ for both species; it vanishes for identical substances, $q_1 = q_2$ and $\upsilon_1 = \upsilon_2$ (the unary nucleation limit). Indeed, $\lambda_{TT} = \lambda_{EE}$ in unary nucleation [17]. The relation between $\langle \Delta x \rangle$ and $\langle \Delta T \rangle$ is shown graphically in Fig. 7.

The same quantities, $\langle \Delta x \rangle$ and $\langle \Delta T \rangle$, for the $H_2O - H_2SO_4$ system are shown in Figs. 5 and 6. As was shown above, the limiting case $\lambda_{xx} \to 0$ takes place here due to the small amount of acid component in the vapor phase and, as a consequence, the nucleation process is isothermal; the function $\psi(\beta_1)$ in Fig. 3 is equal to unity. The calculated dependence $\langle \Delta T(\beta_1) \rangle$ confirms this fact: $\langle \Delta T \rangle \sim 10^{-4} - 10^{-3}$ up to $\beta_1 = 0.9$ and only at $\beta_1$-values close to unity it becomes percentages of the degree, since the $\lambda_{EE}$-value becomes small and nonisothermal effect begins to manifest itself. The isothermal character of the process is also confirmed by the calculated angles $\theta_1 = 1.026^\circ$, $\theta_2 = 88.974^\circ$, and $\theta_3 = 89.999^\circ$ for $\beta_1 = 0.1$. Thus the flux vector lies in the $(V,x)$-plane, i.e. the process is described by the $2D$ $(V,x)$-theory; the obtained $\theta_1$-value is close to the limiting flux direction (Eq. (104)) $\theta_1^{\lim} = 1.031^\circ$ for $\beta_1 = 0.1$. Since this property of the nucleation process is only due to the small fraction of acid in the vapor, it will remain in more realistic (beyond a regular solution) models also.

The mean enrichment $\langle \Delta x(\beta_1) \rangle$ is positive here and has about the same values as in the previous *o*-xylene-*m*-xylene example. The values of $\langle \Delta x(\beta_1) \rangle$ would seem to be greater than in the previous case in view of the smallness of $\lambda_{xx}$ (which means the slow kinetics of composition change). However, we have the large value of $h_{xx}$ here, $h_{22} \approx 2.3 \times 10^3$ versus $h_{22} \approx 179$ for the previous case. Thus, large deviations of the composition from $x_*$ are "forbidden" by thermodynamics. Also $\langle \Delta x(\beta_1) \rangle$ is practically constant in the wide range in $\beta_1$ except the vicinity of $\beta_1 = 1$, where it decreases and even changes the sign. In whole, the behavior of the dependences $\langle \Delta x(\beta_1) \rangle$ and $\langle \Delta T(\beta_1) \rangle$ is similar.

## 5. Summary and conclusions

The single-component $(V,T)$-theory is naturally extended to a binary case by introducing the additional stable variable – droplet composition $x$. The work $W$ of droplet formation in the $(V,x,T)$-theory is a quadratic form with diagonal matrix. Macroscopic growth equations of the droplet underlie the presented approach. They include the equilibrium partial pressures $P_{ei}$ of each species for the droplet of solution of an arbitrary radius and temperature; these pressures in the single-component case are given by Kelvin's equation. The problem of deriving the equations for these pressures in a binary case is presented here as a problem of solving of a Pfaffian equation. It is shown that an explicit generalization of Kelvin's equation to a binary case can be done if either the droplet surface tension or the partial molecular volume does not depend on composition; the required equations are obtained for both these cases. Equations for $P_{ei}$ also serve for deriving different relations between thermodynamic parameters of the droplet and the vapor being in equilibrium. In this way, such an important relation as the dependence of the critical droplet composition on its radius is obtained, as well as the critical radius itself is calculated for the given vapor state.

The droplet motion equations in the $(V,x,T)$-space – equations for $\dot{V}$, $\dot{x}$, and $\dot{T}$ - in the near-critical region are written on the basis of the previously developed algorithm [25] with the help of the macroscopic growth equations mentioned above. The characteristic kinetic parameters governing the nucleation kinetics are determined with the help of this algorithm; Onsager's reciprocal relations are employed for calculating these parameters as well. The equation for $\dot{T}$ obtained in this way is an energy balance equation, or the first law of thermodynamics for the droplet, which speaks in favor of the self-consistency of the theory. One more feature of the self-consistency is given by transition from the $(V,x,T)$- to $(N_1,N_2,E)$-variables; the meaning of the equation for $\dot{T}$ is confirmed as well as the derived equation for the diffusion tensor is a clear extension of the corresponding single-component equation. The matrix $\mathbf{Z}$ of the motion equations determines the droplet flux direction in the vicinity of the saddle point of the $3D$ hypersurface $W(V,x,T)$; the nucleation barrier curvature in this direction is given by Euler's formula. The matrix $\mathbf{Z}$ negative eigenvalue $\kappa_1$ is included by the nucleation rate and therefore determines the nucleation kinetics. The limiting effect of kinetic processes on nucleation is shown by considering different limiting cases with respect to the kinetic parameters. Nonisothermal effect decreasing the nucleation rate in comparison with the isothermal one is also the manifestation of the limiting character of the kinetic process – the heat exchange between the droplet and the vapor.

As an example for demonstrative numerical calculations, the two systems are employed: ideal *o*-xylene-*m*-xylene and regular water-sulfuric acid systems. Though the latter solution is not described quantitatively by the regular solution model, some qualitative features of the nucleation kinetics can be understood within this simple model; they are due to the two main characteristics of the acid – very small saturation pressure and large mixing heat. The calculations show that the nucleation kinetics in the *o*-xylene- *m*-xylene vapor is qualitatively the same as in the unary theory; the difference only in that the enrichment effect appears alongside with the nonisothermal effect. Thus, both the mean overheat of droplets and their mean enrichment exist in the steady state. These quantities are related to each other, since the droplet growth equations include both the temperature and composition via the partial equilibrium pressures $P_{ei}$.

The nucleation kinetics in the water-sulfuric acid vapor essentially differs from that in the *o*-xylene-*m*-xylene system. Here there is one small kinetic parameter $\lambda_{xx}$ due to the small amount of the acid in the vapor. As a consequence, (i) the parameter $\kappa_1$ and the nucleation rate are proportional to $\lambda_{xx}$ and (ii) the nucleation process is isothermal in the almost whole range of the condensation coefficient values except the vicinity of $\beta_1 = 1$; the mean steady-state enrichment is constant in the same interval. This result allows using the isothermal theory for studying the nucleation kinetics in the given system with more realistic models.

## References

[1] G. J. Doyle, J. Chem. Phys. 35 (3) (1961) 795-799.

[2] P. Mirabel, J. L. Katz, J. Chem. Phys. 60 (3) (1974) 1138-1144.

[3] H. Reiss, J. Chem. Phys. 18 (1950) 840-848.

[4] Ya. B. Zeldovich, J. Exp. Theor. Phys. 12 (1942) 525-538.

[5] J. Frenkel, *Kinetic Theory of Liquids* (Oxford, New York, 1946).

[6] D. Stauffer, J. Aerosol Sci., 7 (1976) 319-333.

[7] N. V. Alekseechkin, P. N. Ostapchuk, Fiz. Tverd. Tela 35 (4) (1993) 929-940 [Phys. Solid State 35 (4) (1993) 479-484; American Institute of Physics].

[8] G. Wilemski, J. Chem. Phys. 110 (13) (1999) 6451-6457.
[9] N. V. Alekseechkin, J. Chem. Phys. 124 (2006) 124512.
[10] H. Reiss, W. K. Kegel, J. Phys. Chem. 100 (1996) 10428-10432.
[11] H. Reiss, W. K. Kegel, J. L. Katz, Phys. Rev. Lett. 78 (1997) 4506-4509.
[12] H. Reiss, W. K. Kegel, J. L. Katz, J. Phys. Chem. A 102 (1998) 8548-8555.
[13] B. E. Wyslouzil, G. Wilemski, J. Chem. Phys. 103 (3) (1995) 1137-1151.
[14] B. E. Wyslouzil, G. Wilemski, J. Chem. Phys. 105 (3) (1996) 1090-1100.
[15] S. P. Fisenko, G. Wilemski, Phys. Rev. E 70 (2004) 056119.
[16] J. Feder, K. C. Russell, J. Lothe, G. M. Pound, Adv. Phys. 15 (1966) 111-178.
[17] N. V. Alekseechkin, Physica A 412 (2014) 186-205.
[18] I. J. Ford, C. F. Clement, J. Phys. A 22 (1989) 4007-4018.
[19] J. C. Barrett, C. F. Clement, I. J. Ford, J. Phys. A 26 (1993) 529-548.
[20] B. E. Wyslouzil, J. H. Seinfeld, J. Chem. Phys. 97 (1992) 2661-2670.
[21] J. C. Barrett, J. Phys. A 27 (1994) 5053-5068.
[22] J. C. Barrett, J. Chem. Phys. 128 (2008) 164519.
[23] J. C. Barrett, J. Chem. Phys. 135 (2011) 096101.
[24] J. Wedekind, D. Reguera, R. Strey, J. Chem. Phys. 127 (2007) 064501.
[25] N. V. Alekseechkin, J. Phys. Chem. B 116 (2012) 9445-9459.
[26] N. V. Alekseechkin, Eur. Phys. J. B 86: 401 (2013).
[27] N. V. Alekseechkin, arXiv: 1211. 1085v4 [physics. chem.-ph].
[28] G. Wilemski, J. Chem. Phys. 80 (3) (1984) 1370-1372.
[29] G. Wilemski, J. Chem. Phys. 88 (8) (1988) 5134-5136.
[30] S*ulfuric Acid Handbook*, edited by K. M. Malin (Khimiya, Moscow, 1971).

| parameters | $P_{si}$, *Torr* | $\zeta_i$ | $\xi_i$, *g/cm s* | $\frac{2\upsilon_i\sigma}{kT_0R_*}$ | $\bar{q}_i$ | $P_0$, *Torr* | $x_0$ | $\sigma$, *erg/cm*$^2$ | $c_V^0$ | $c_V$ | $u_1$, *cm/s* |
|---|---|---|---|---|---|---|---|---|---|---|---|
| *o*-xylene | 4.88 | 2.92×10$^3$ | 0.040 $\beta$ | 2.24 | 15.13 | 51.8 | 0.5 | 29 | 14 $k$ | 22 $k$ | 6.05×10$^3$ |
| *m*-xylene | 6.14 | 2.87×10$^3$ | 0.041 $\beta$ | 2.28 | 14.8 | | | | | | |

| $R_*$, *nm* | $x_*$ | $N_*$ | $\lvert z_{VV}\rvert$, $s^{-1}$ | $\lambda_{xx}$, $s^{-1}$ | $\lambda_{TT}$, $s^{-1}$ | $I$, *cm*$^{-3}$ *s*$^{-1}$ | $\lvert h_{11}\rvert$ | $h_{22}$ | $h_{33}$ | $W'_*$ |
|---|---|---|---|---|---|---|---|---|---|---|
| 1.285 | 0.568 | 44 | 3.67×10$^6$ | 4.9×10$^6$ | 2.9×10$^7$ | 6.4×10$^6$ | 33.0 | 179.2 | 967.1 | 49.5 |

Table 1. Physical properties of the *o*-xylene-*m*-xylene system at the temperature $T_0 = 20\ ^\circ C$ together with calculated nucleation parameters (kinetic parameters are given for $\beta = 0.1$).

| parameters | $P_{si}$, *Torr* | $A_{0i}$ | $P_{0i}$, *Torr* | $\rho_{0i}$, *cm*$^{-3}$ | $\xi_i$, *g/cm s* | $\bar{q}_i$ | $c_{Vi}^0$ | $c_V/k$ | $\omega$ |
|---|---|---|---|---|---|---|---|---|---|
| water | 23.78 | 2.7×10$^{-2}$ | 0.64 | 2.07×10$^{16}$ | 8.2×10$^{-5}$ $\beta_1$ | 20.5 | 3.2 $k$ | 12.3 | 25 |
| sulfuric acid | 5×10$^{-4}$ | 2.7×10$^{-3}$ | 1.35×10$^{-6}$ | 4.4×10$^{10}$ | 6.4×10$^{-10}$ $\beta_2$ | 24.7 | 8.7 $k$ | | |

| $R_*$, *nm* | $x_*$ | $N_*$ | $\lvert z_{VV}\rvert$, $s^{-1}$ | $\lambda_{xx}$, $s^{-1}$ | $\lambda_{TT}$, $s^{-1}$ | $I$, *cm*$^{-3}$ *s*$^{-1}$ | $\lvert h_{11}\rvert$ | $h_{22}$ | $h_{33}$ | $\sigma = \sigma(x_*)$, *erg/cm*$^2$ | $W'_*$ |
|---|---|---|---|---|---|---|---|---|---|---|---|
| 0.817 | 0.586 | 42 | 1.4×10$^4$ | 48.6 | 1.7×10$^5$ | 1.6×10$^3$ | 30.5 | 2.3×10$^3$ | 516.6 | 67.46 | 45.8 |

Table 2. Physical properties of the $H_2O - H_2SO_4$ system at the temperature $T_0 = 25\ ^\circ C$ together with calculated nucleation parameters (kinetic parameters are given for $\beta_1 = 0.1$).

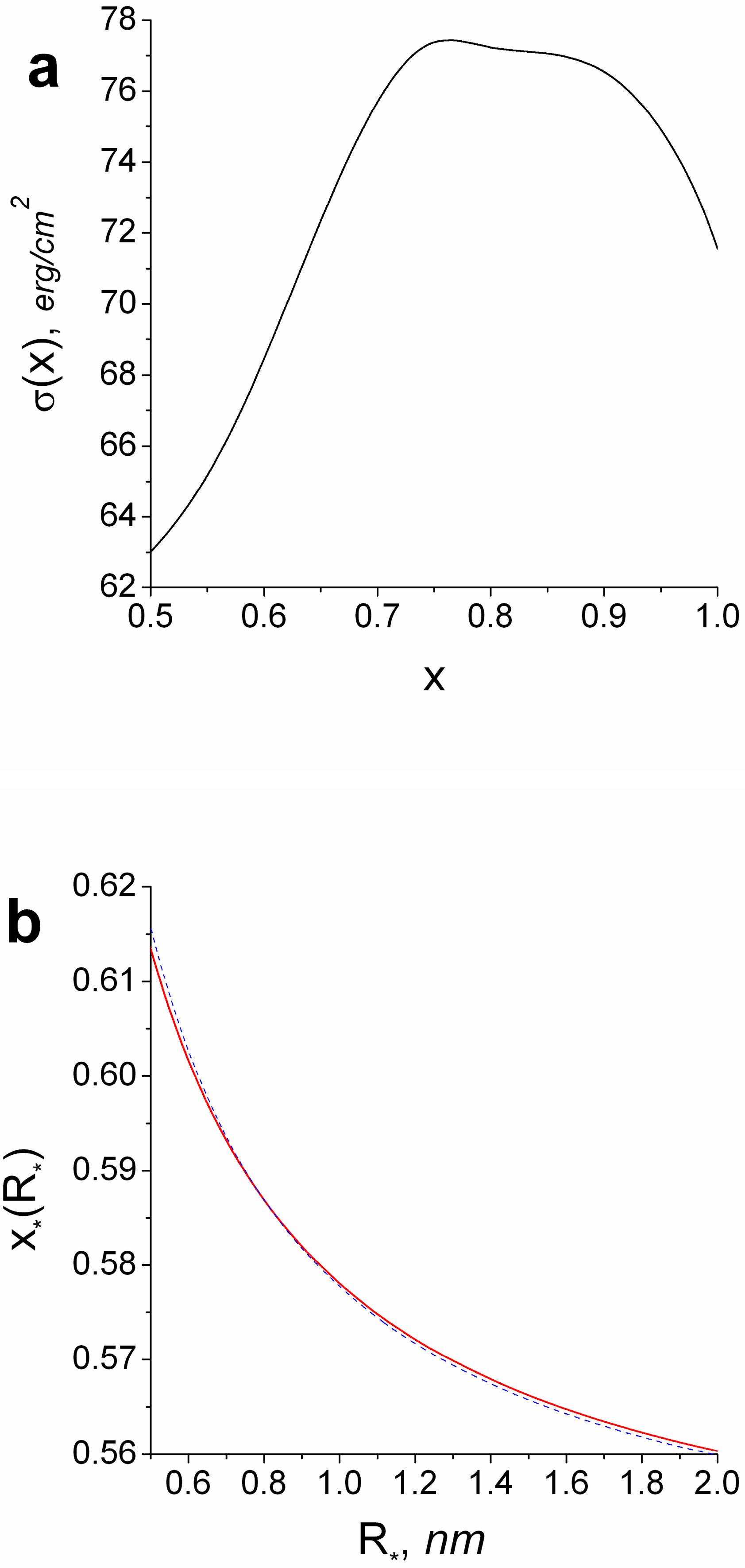

Fig.1. (a) Cubic-spline dependence of the $H_2O - H_2SO_4$-solution surface tension on composition at $20\ ^{\circ}C$ built by experimental points [30]. (b) Dependence of the critical composition on radius, Eq. (49), for the constant surface tension $\sigma = \sigma(0.586) = 67.46$ (solid) and the composition-dependent surface tension given in Fig. (a) (dashed).

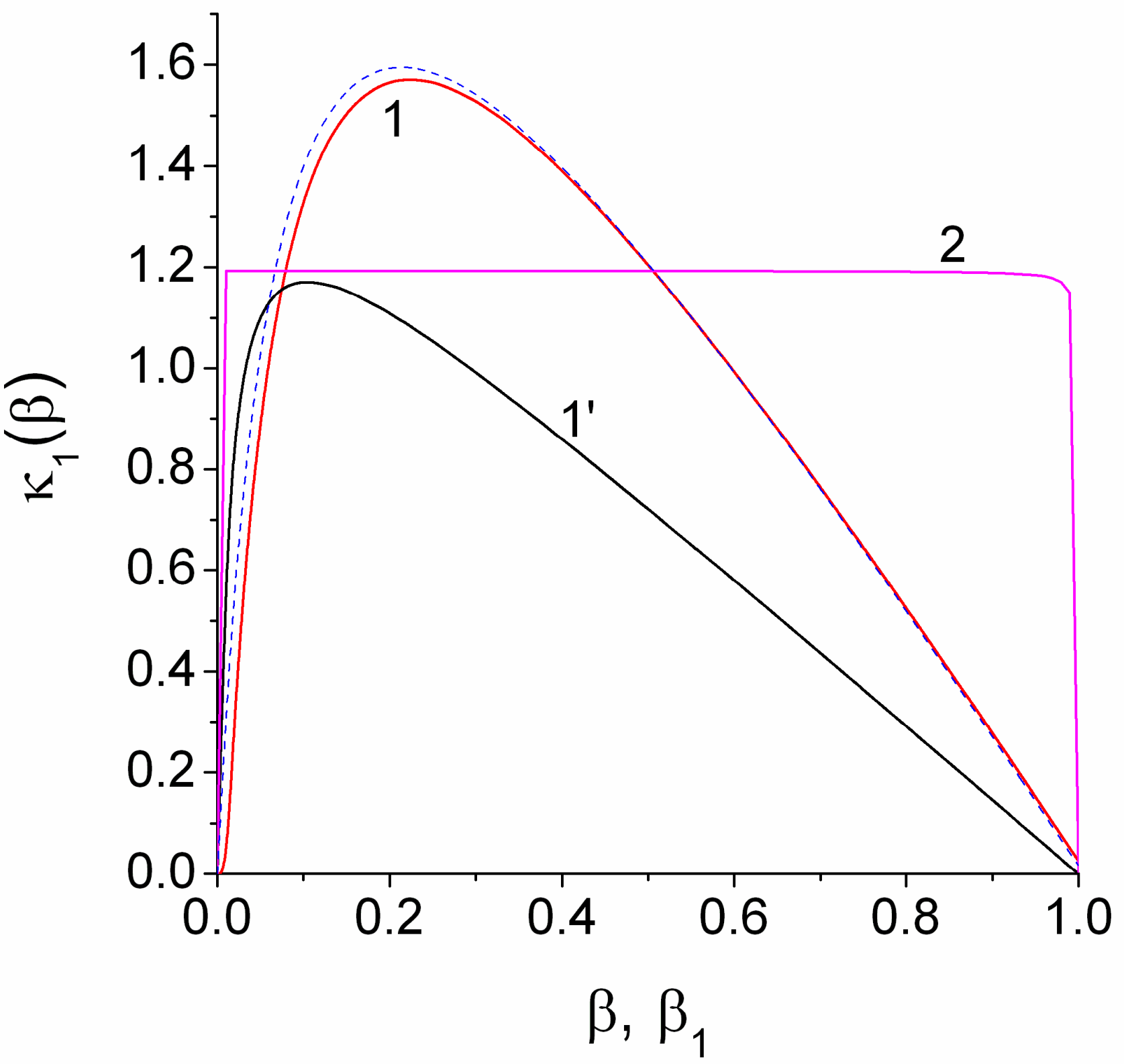

Fig. 2. Dependences $\kappa_1(\beta)\times10^{-6}$, $s^{-1}$, for the *o*-xylene-*m*-xylene system (solid 1); $\kappa_x(\beta)\times10^{-6}$ in the unary limit, solution of Eq. (96) – dashed; $\kappa_1(\beta_1)$ for the $H_2O - H_2SO_4$ system (solid 2), and $\kappa_1(\beta)\times10^{-5}$ for the unary nucleation in water vapor studied in the previous work ($1'$).

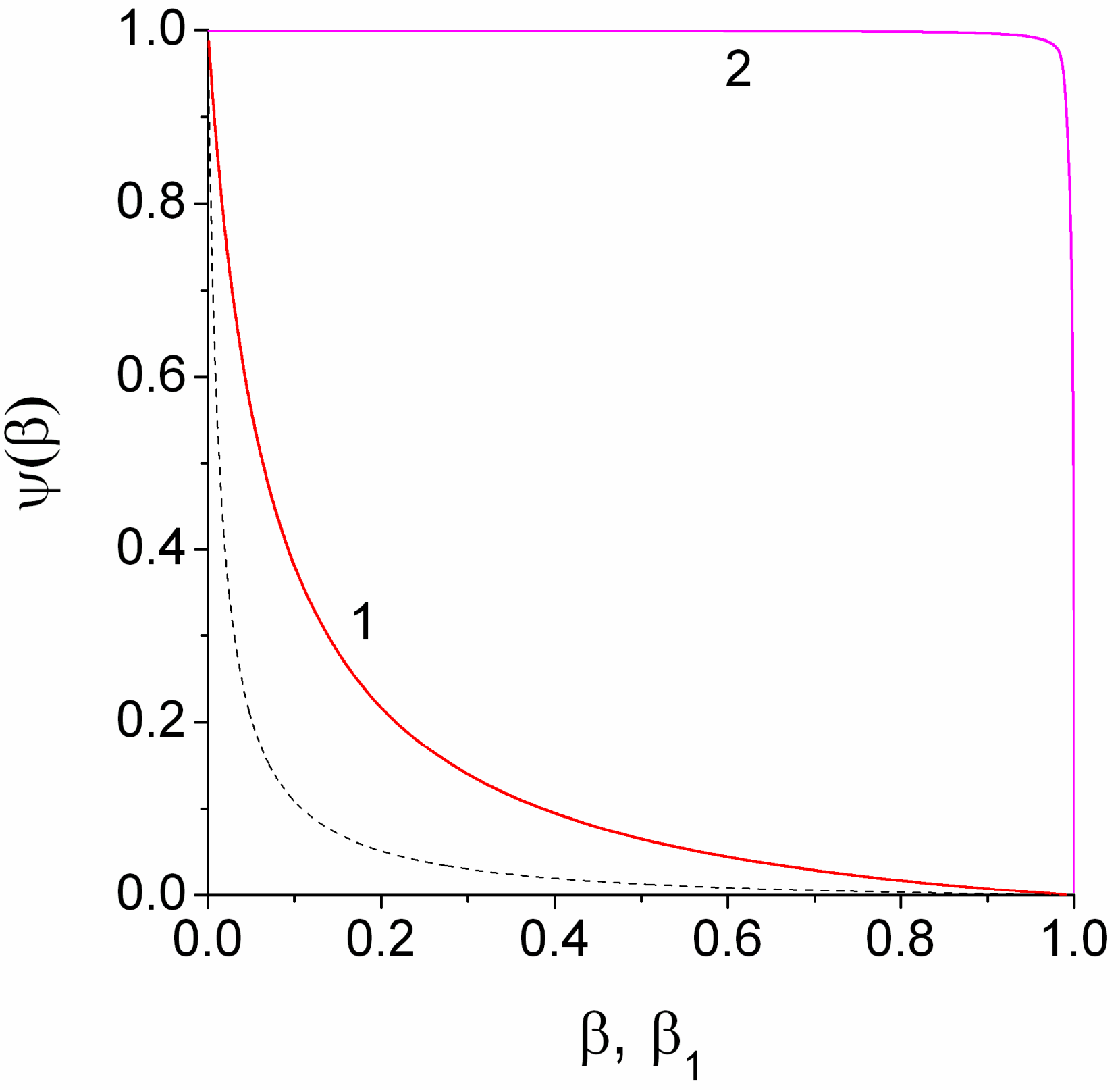

Fig. 3. Nonisothermal effect $\psi(\beta) = I / I_{iso}$ for the *o*-xylene-*m*-xylene system (1), $\psi(\beta_1)$ for the $H_2O - H_2SO_4$ system (2), and for the unary nucleation in water vapor (dashed).

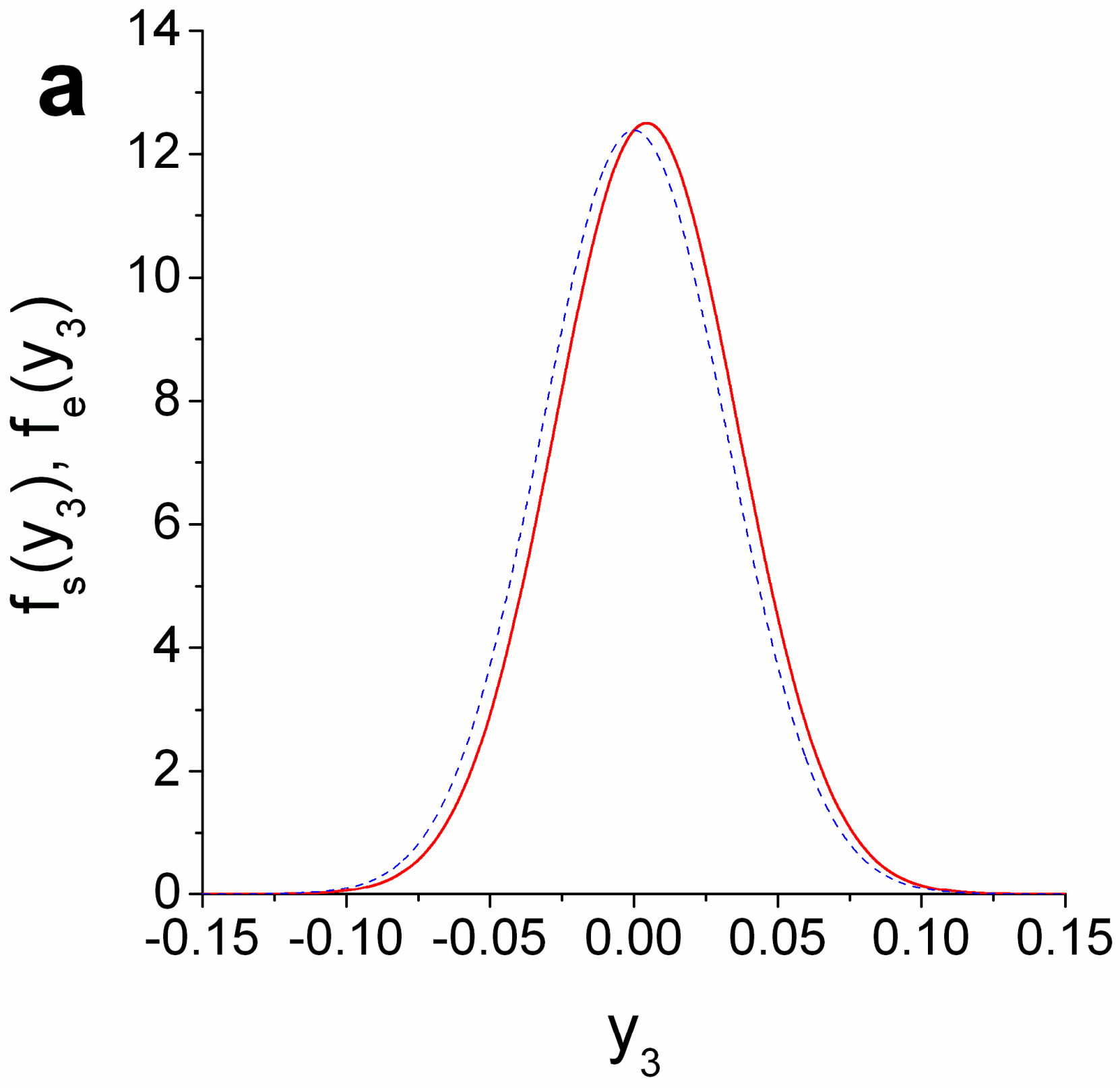

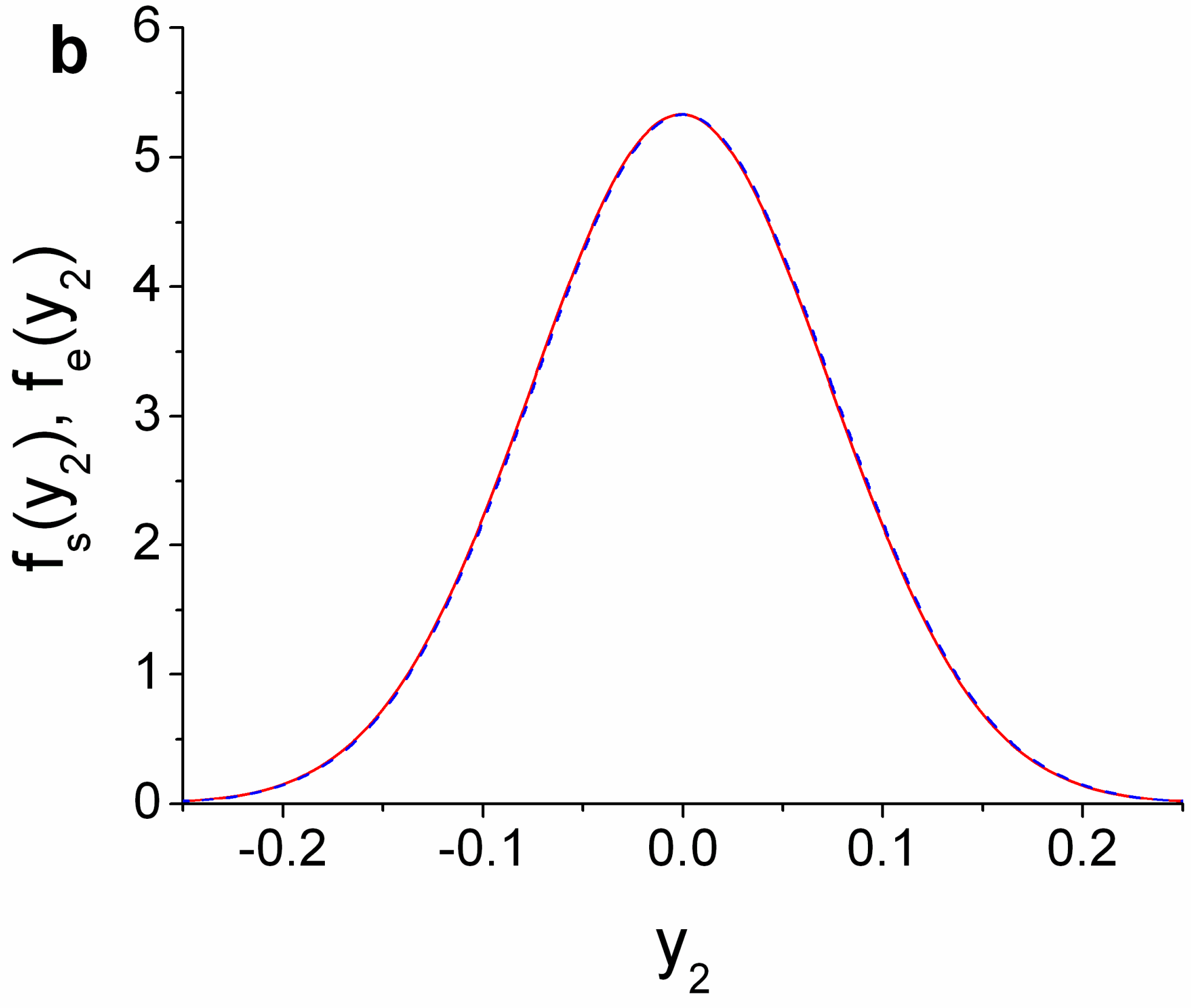

Fig. 4. Steady state (solid) and equilibrium (dashed) distributions of droplets on temperature (a) and composition (b), Eqs. (112).

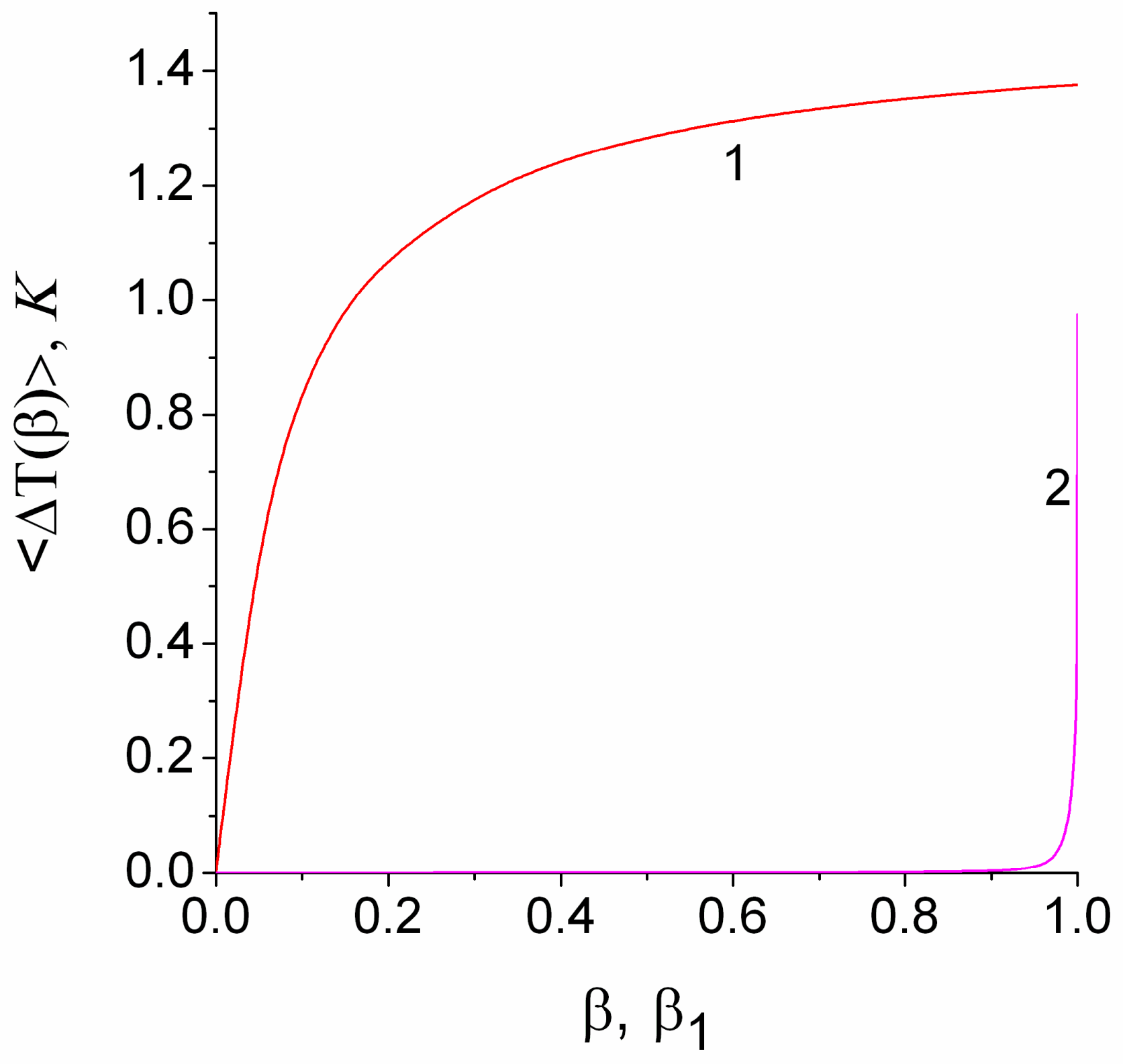

Fig. 5. Steady-state mean overheat of droplets in the near-critical region $-\delta < y_1 < \delta$, $\delta = 2.5$, for the *o*-xylene-*m*-xylene system (1) and for the $H_2O - H_2SO_4$ system (2) as a function of the condensation coefficient.

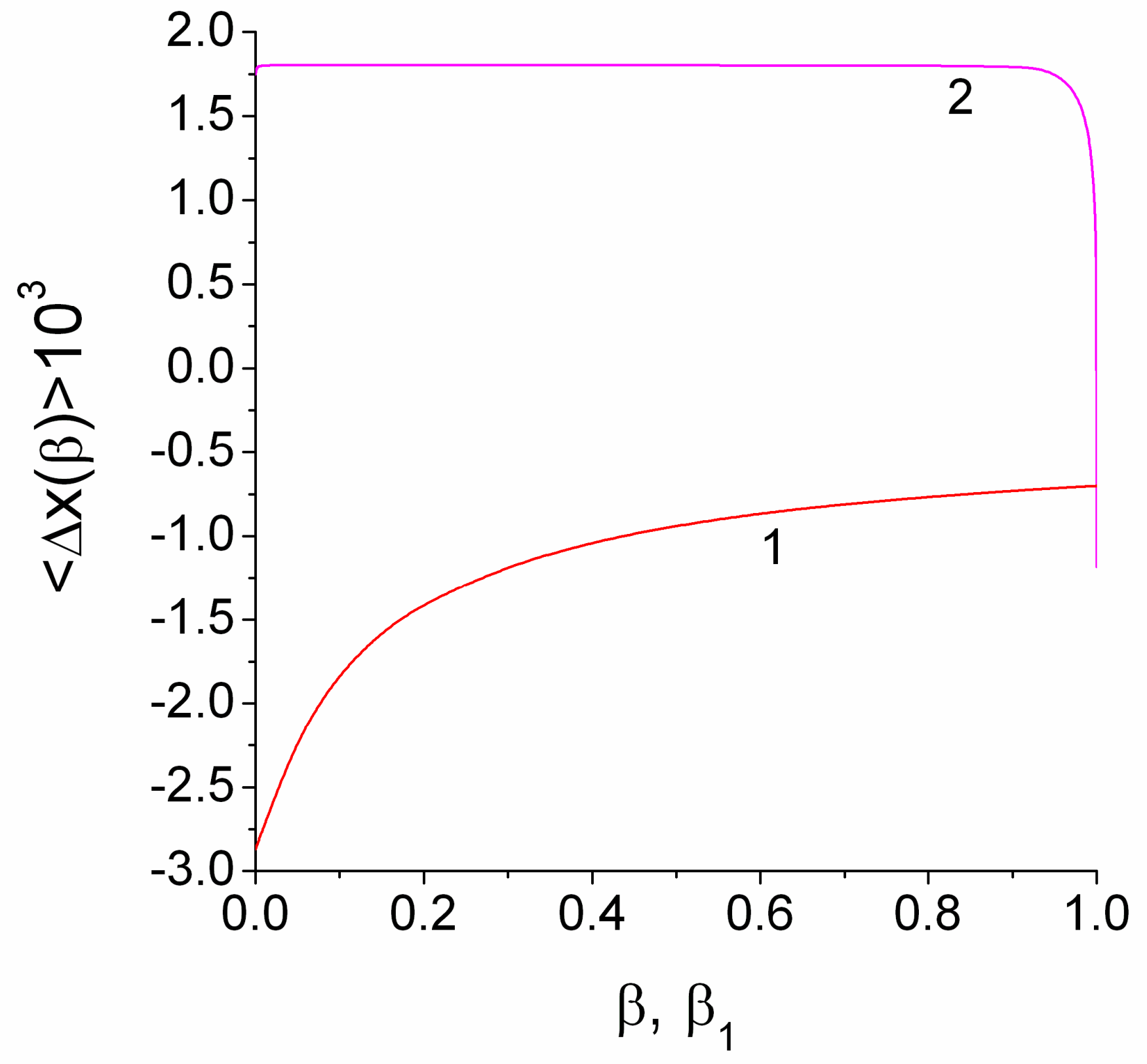

Fig. 6. Steady-state mean enrichment of droplets in the near-critical region, for the *o*-xylene-*m*-xylene system (1) and for the $H_2O-H_2SO_4$ system (2) as a function of the condensation coefficient.

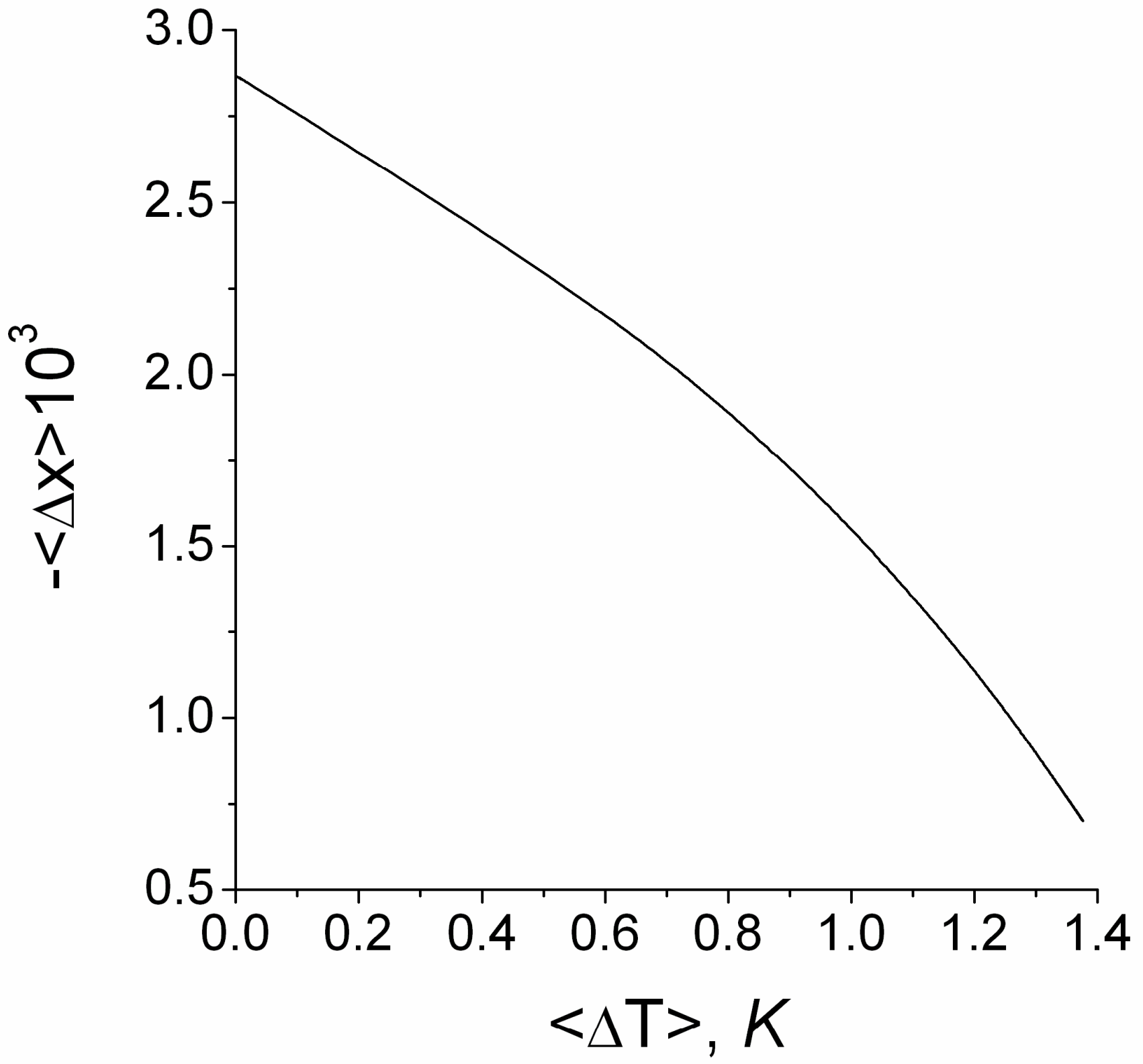

Fig. 7. Absolute value of the mean enrichment versus the mean overheat for the *o*-xylene-*m*-xylene system.